\documentclass{rstransa}

\usepackage{graphicx}
\usepackage{amsmath}
\numberwithin{equation}{section}
\usepackage{amssymb}
\usepackage{mathtools}

\usepackage{hyperref}

\usepackage{amsthm}
\newtheorem{theorem}{Theorem}
\newtheorem{remark}{Remark}

\newcommand{\vect}[1]{\boldsymbol{#1}}
\def\NS{Navier-Stokes}
\def\YM{Yang-Mills}
\def\VEV#1{\left\langle #1 \right\rangle}
\def\I{\mathrm{i}}

\def\tr{\mathrm{tr}}
\def\pd{\partial}
\def\ff#1{\frac{\delta}{\delta #1}}
\def\fbyf#1#2{\frac{\delta #1}{\delta #2}}
\DeclarePairedDelimiter{\floor}{\lfloor}{\rfloor}
\def\Ra{\Rightarrow}
\def\lb#1{\left[#1\right.}
\def\rb#1{\left.#1\right]}

\newcommand{\Mathematica}{\textit{Mathematica\textsuperscript{\resizebox{!}{0.8ex}{\textregistered}}}}
\def\8{\infty}

\def\const{\textit{ const }}
\def\undertext#1{\vtop{\hbox{#1}\kern 1pt \hrule}}
\def\Ra{\Rightarrow}

\def\abs#1{\left| #1\right|}

\def\pd#1{\partial_{#1}}
\def\VEV#1{\left\langle #1\right\rangle}

\def\tr{\hbox{tr}\,}

\def\ff#1{\frac{\delta}{\delta#1}}
\def\fbyf#1#2{\frac{\delta#1}{\delta#2}}

\def\bea{\begin{eqnarray} && &&}
\def\eea{\end{eqnarray}}

\let\oldexp\exp
\renewcommand{\exp}[1]{\oldexp\left(#1\right)}

\def\NS{Navier-Stokes}
\def\YM{Yang-Mills}

\newcommand{\Mod}[1]{\ (\mathrm{mod}\ #1)}

\def\XXint#1#2#3{{\setbox0=\hbox{$#1{#2#3}{\int}$}
     \vcenter{\hbox{$#2#3$}}\kern-.5\wd0}}

\renewcommand{\Re}{\textbf{Re }}

\newcommand{\tpmod}[1]{{\@displayfalse\Mod{#1}}}
\usepackage[english]{babel}

\def\loopcalculus{loop space calculus }

\begin{document}

\title{Geometric Solution of Turbulence as Diffusion in Loop Space}

\author{Alexander Migdal}

\address{Institute for Advanced Study, Princeton, NJ, USA}

\subject{Fluid dynamics, Statistical physics, Theoretical physics}

\keywords{turbulence, Navier-Stokes, MHD, Yang-Mills theory, loop equations, string theory duality, Riemann zeta function}

\corres{Alexander Migdal\\\email{amigdal@ias.edu}}

\begin{abstract}
Strongly nonlinear dynamics, from fluid turbulence to quantum chromodynamics, have long constituted some of the most challenging problems in theoretical physics. This review describes a unified theoretical framework, the loop space calculus, which offers an analytical approach to these problems. The central idea is a shift in perspective from pointwise fields to integrated loop observables, a transformation that recasts the governing nonlinear equations into a universal linear diffusion equation in the space of loops. This framework, supported by recent mathematical analysis, is analytically solvable and yields an exact, parameter-free solution for decaying hydrodynamic turbulence—the Euler ensemble—which is shown to be dual to a solvable string theory. The theory's predictions include: (i) the unification of spatial and temporal scaling laws, governed by two related, infinite spectra of intermittency and decay exponents derived from the nontrivial zeros of the Riemann zeta function; (ii) a first-order phase transition in magnetohydrodynamic (MHD) turbulence; and (iii) the formation of quantized, concentric shells in passive scalar mixing.  The appearance of identical mathematical structures as solutions to the turbulent regime of Yang--Mills gradient flow points to the broad applicability of this approach. The framework also yields a new type of \emph{analytic Hodge-dual matrix surface} that solves the Yang--Mills fixed-point loop equation by harmonic map, opening the way for a geometric formulation of QCD string theory.
\end{abstract}

\maketitle

\section{Introduction}

In his 1964 \textit{Lectures} \cite{Feynman}, Richard Feynman identified the analysis of turbulent fluids as a problem "left over from a long time ago... Nobody in physics has really been able to analyze it mathematically satisfactorily." Decades later, this challenge, along with the related problem of non-perturbative dynamics in Yang-Mills gauge theory, has remained a frontier of theoretical science. These phenomena are governed by strongly nonlinear equations that defy standard analytical techniques, forcing a reliance on phenomenological models or numerical simulation.

An analytical approach to these problems emerges from a new theoretical paradigm, which is the subject of this review. The core idea is to shift the descriptive language of physics away from local, pointwise fields (like velocity $\vect{v}(\vect{r})$ or the gauge potential $A_\mu(x)$) and towards non-local, integrated observables defined on closed loops (like velocity circulation $\Gamma_C$ or the Wilson loop $W[C]$). 

This reformulation is not merely a change of variables; it constitutes a dimensional reduction of the dynamics that also eliminates the nonlinearity. It transforms the governing partial differential equations into a universal, linear diffusion equation in the space of loops, rendering the statistical evolution analytically tractable via a functional Fourier transformation to momentum loop space, where the loop-space diffusion reduces to an algebraic problem. This evolution converges to a universal attractor — the Euler ensemble — which is dual to a solvable string theory.

A central result of this framework is its prediction of two distinct but related spectra of exponents. For the spatial scaling of velocity correlations, where phenomenological models typically assume a single dominant power law for each statistical moment, the theory reveals an infinite, discrete spectrum of \textbf{intermittency exponents}. The leading exponent in this series, which governs the dominant large-scale behavior, can be regarded as the "ground state" of the scaling law. The subsequent exponents then define a hierarchy of corrections to this leading behavior, akin to quantum corrections, which become relevant at finer scales. In the same manner, the theory determines a full spectrum of \textbf{decay exponents} for the temporal evolution of integrated quantities, such as the kinetic energy. Here too, the principal exponent, which corresponds to the observed decay law ($E \propto t^{-5/4}$), acts as the ground state, with the rest of the spectrum describing higher-order corrections to the temporal decay. Crucially, these two spectra, traditionally considered independent, emerge from the same underlying number-theoretic structure of the Euler ensemble. This solution establishes a direct connection between the fine-scale spatial statistics and the long-term temporal decay of the turbulent flow.

The Euler ensemble solution provides an explicit analytic expression for the full spectrum of these exponents, relating them to rational numbers and the complex zeros of the Riemann zeta function. The imaginary parts of the complex exponents predict oscillations in correlation functions on a logarithmic scale.

\subsection*{A Brief Historical Context}

The challenge of deriving a statistical theory of turbulence directly from the Navier-Stokes equations has a long history. Notably, Eberhard Hopf pioneered the use of functional methods, introducing an equation for the characteristic functional of the velocity field and envisioning a universal turbulent attractor governing the long-time statistics (see a recent review  \cite{Hopf19}). However, Hopf's functional equation proved intractable, leaving the nature of the attractor unclear for decades. Now, with the loop equation replacing the Hopf functional equation, we have its solution —the Euler ensemble —as the realization of Hopf's conjecture about the turbulent attractor.

Later, the loop space formalism emerged, rooted in the Makeenko-Migdal (MM) equations of the late 1970s and early 1980s~\cite{MMEq79, MM1981NPB, Mig83}. These were immediately recognized in the high-energy physics community as the first exact, non-perturbative dynamical equations for QCD. However, the initial hope of finding an analytical solution for quark confinement soon faded. The mathematical formulation, while geometrically insightful, proved analytically intractable, hampering further progress and shifting the community's focus toward lattice gauge theory. The loop approach offered a different path compared to Hopf's formulation, focusing on circulation observables.

The scope of the formalism was extended to fluid dynamics in the early nineties with the derivation of the loop equations for the Navier-Stokes system~\cite{M93}. Within this framework, a WKB solution was derived for the limit of large, smooth loops. This solution established a theoretical "area law" for the statistics of velocity circulation. A transformation to a dual "momentum loop space" was also shown to reduce the problem to a singular, one-dimensional equation. For decades, however, further progress was impeded by the mathematical challenges posed by this equation and by a lack of high-precision data with which to test the theory's predictions.

The theoretical landscape evolved considerably after 2019, when high-resolution direct numerical simulations (DNS) by Sreenivasan and collaborators reported results consistent with the area law derived nearly three decades earlier~\cite{S21, Pumir2021Minimal}. 
This numerical evidence renewed interest in the approach. At the phenomenological level, the statistics of velocity circulation, the Area law, and its various applications to classical and quantum turbulence were discussed in a series of important papers, elucidating both the microscopic meaning, numerical verification by DNS and the applicability of the circulation statistics to the turbulence problem \cite{Polanco2021NatComm,Muller2021PRX,Moriconi2022PRE,Zhu2023PRL,Iyer2024PhysFluids,Moriconi2025arXiv}.

This development also provided the impetus to resolve the long-standing mathematical challenges of the momentum loop equation.
The complete analytical solution that emerged, realizing aspects of Hopf's original vision via loop space, is the subject of this review. (See also \cite{migdal2024quantum} for further discussion on the connection to Hopf's work).

This history underscores a critical point: a correct, nonsingular, and tractable loop space formalism was the missing key. This review introduces this key---a new, nonsingular calculus that renders the loop equations solvable.

\subsection*{Recent Mathematical and Numerical Validation}
This theoretical structure, initially derived from physical principles, has recently received significant independent validation. From the number theory perspective, the statistical properties of the Euler ensemble at finite $N$ were studied in \cite{Zah23} by combinatorial methods, confirming and extending the results and conjectures made in the first paper \cite{migdal2023exact}. In addition, the distribution of radii of regular star polygons with unit side lengths was recently reproduced by Debmalya Basak using rigorous methods of number theory (private communication, to be published).
From a \NS{} dynamics perspective, a rigorous analysis by Brue and De Lellis \cite{DeLellisInprep} has confirmed that the Euler ensemble is an exact solution to the discrete version of the loop equation (Theorem 10.6 in their paper). While the final step---proving the existence of the continuum limit of the fluid dynamics observables as the number of loop segments $N \to \infty$---remains an open mathematical problem, their result provides a firm mathematical foundation for the core of the theory. It confirms that the Euler ensemble is not an approximation but an exact solution at the level of the regularized theory.
\begin{remark}
Two other issues were left unresolved in that paper :(i) Proving the global uniqueness of the probability measure on the (loop) space of trajectories for appropriately chosen ensemble of initial conditions, and (ii) The boundedness of error arising from the discretization of the loop equation with or without the “liquid loop approach”. Global uniqueness may not even be true: the fixed points of nonlinear systems like \NS{} equations, as we know from the general RG theory of critical phenomena, are not unique, so there could be other fixed points in addition to the Euler ensemble. The empirical evidence from real and numerical experiments (below) speaks in favor of such universality of the Euler ensemble. The boundedness of error, in our opinion, is just a temporary issue, soon to be resolved. We developed another approach to the polygonal estimates by Brue and de Lellis, which we plan to develop as a rigorous theorem in collaboration with Google AI lab.
\end{remark}
This mathematical validation is now strongly complemented by \textbf{new high-resolution $4096^3$ DNS results from Sreenivasan and collaborators} \cite{SreeniAkash2025}. These results \textbf{verify the theory's quantitative predictions with high precision}, representing a significant development. As discussed in Sections 4.1 and 8, the DNS data for key observables—such as the $E \propto t^{-5/4}$ decay law and the universal, nonlinear effective index—are in \textbf{excellent, parameter-free agreement} with the theoretical curves derived from the Euler ensemble (see Fig. 2).
\begin{remark}
    Initial results from \cite{SreeniAkash2025} seemed to contradict predictions of the Euler ensemble, but this confusion was cleared in discussions with these authors and is reflected both in their final paper , submitted to this issue, and in this paper below. The issue of apparent violation of universality was that unlike the effective scaling index $\xi_2(r,t)$, which is dominated by the bulk part of the decaying energy spectrum, the decay of total energy is sensitive to the small wave-vectors, related to the large scale boundary conditions, which are clearly non-universal. The Euler ensemble was never intended to describe such boundary effects; it was built as a theory of homogeneous turbulence in infinite space. In the particular case of the $4096^3$ DNS, these large scale boundary effects are polluted by lattice artifacts ( the integer wave-vectors $k=0,1,2...$ in inverse lattice units). The authors of \cite{SreeniAkash2025} investigated the influence of these artifacts by cutting out several first integer values of $k$ in the energy spectrum. The total energy was indeed influenced, and that explained most of the non-universality they observed in energy decay, fitted by K41 scaling laws. The enstrophy is less sensitive to these artifacts, and it matches the Euler ensemble with better accuracy. Finally, there are non-universality corrections from subleading power decay terms, as predicted by the Euler ensemble. These corrections may lead to systematic errors $~10\%$ on top of the lattice artifacts.
\end{remark} 
Together, these mathematical and numerical results provide strong support for the loop space calculus in the Navier-Stokes equation, confirming the Euler ensemble as a robust solution for decaying turbulence.
\subsection*{Outline of the Review}
This review is organized as follows. The discussion begins with an introduction to the general loop space approach, applicable to both abelian fields in fluid dynamics and nonabelian fields in Yang-Mills theory. The approach is then applied to a hierarchy of turbulence problems, beginning with the foundational case of decaying hydrodynamic turbulence, then the transport of passive scalars, and finally the more complex case of magnetohydrodynamic (MHD) turbulence. A subsequent section situates these analytical solutions within the field's historical context and clarifies their relationship to classical paradigms of turbulence theory. 
A particularly striking consequence of this framework emerges in gauge theory.
In the Yang--Mills gradient flow, the same loop-space formalism leads to an
exact analytic solution---the \emph{Hodge-dual matrix surface}---which solves
the fixed-point Yang--Mills loop equation by harmonic map \cite{migdal2025geometric}.
This result establishes a concrete geometric foundation for the QCD confining
string and provides a direct analytical bridge between turbulence, gauge theory,
and string dualities.

\section{The Loop Space Approach: From Nonlinear PDEs to Linear Diffusion}

The foundation of our approach is the loop functional, $\Psi[C]$, which acts as` the characteristic function for the probability distribution of a loop observable. For fluid dynamics, this is the circulation, $\Gamma_C = \oint_C \vect{v} \cdot d\vect{r}$, and the functional is $\Psi[C] = \langle \exp{\I \Gamma_C / \nu} \rangle$. For gauge theory, it is the Wilson loop.

\subsection*{\NS{} circulation equation}

Let us start with the \NS{} equation for an incompressible velocity field
\begin{align}
    \pd{t} v_\alpha = - v_\beta\pd{\beta} v_\alpha + \nu \pd{\beta} \pd{\beta} v_\alpha - \pd{\alpha} p;\; \pd{\beta}v_\beta =0;
\end{align}
\begin{remark}
    The boundary conditions are vanishing velocity at infinity (or arbitrary constant velocity, which is equivalent by virtue of Galilean invariance). The constant velocity ( or any potential velocity $v_\mu = \pd{\mu} \phi$ ) drops  from the circulation we are investigating in this paper.
\end{remark}
This equation leads to the well-known expression for the time derivative of the circulation, which we write in geometric form using the covariant derivative operator 
\begin{equation}
    D_\mu = \pd{\mu} + \frac{\I v_\mu}{\nu};
\end{equation}
Here is the covariant equation for the time derivative of the circulation
\begin{align}\label{staticEq}
    &\pd{t} \Gamma = \nu \oint_C d x_\alpha\left( \left[D_\beta ,\omega_{\beta\alpha}\right] + v_\beta\omega_{\beta\alpha}\right);\\
    & \omega_{\mu\nu} = \pd{\mu} v_\nu -\pd{\nu} v_\mu ; 
\end{align}
The covariant derivative operator $D_\mu$ will play an important role in the loop dynamics. Note that the imaginary unit in front of the velocity field is similar to that in the Abelian gauge theory covariant derivative, keeping the covariant derivative operator anti-Hermitian $ D_\mu^\dag = - D_\mu$.
The vorticity itself can be represented as a commutator.
\begin{align}
   \I \omega_{\mu\nu}   =\nu [D_\mu, D_\nu];
\end{align}
The last term in \eqref{staticEq} can be eliminated by switching to the "liquid loop" - the one with each point moving with the local velocity. For such a loop, the Kelvin theorem states that the advection term $v_\beta\omega_{\beta\alpha}$ is exactly canceled by the term coming from the loop motion. 
\begin{align}
    \pd{t} C_\alpha(\theta) = v_\alpha(C(\theta), t)
\end{align}
As a result, the circulation $\tilde \Gamma$ of the moving loop satisfies a simple geometric equation
\begin{align}
    &\I\pd{t} \tilde \Gamma = \nu^2 \oint_C d x_\alpha \left[D_\beta ,[D_{\beta}, D_{\alpha}]\right]
\end{align}
after which the \NS{} equation for liquid loop functional $\tilde \Psi$ takes a purely geometric form
\begin{align}
 & \pd{t} \tilde \Psi[C] = \nu\oint_C d x_\alpha \left[D_\beta ,[D_{\beta}, D_{\alpha}]\right] \tilde \Psi[C];
\end{align}
\begin{remark}
    The loop $C$ in our functional needs to be continuous for gauge invariance in the \YM{} case and for the independence of the potential part of velocity in the \NS{} case. No smoothness is assumed nor needed: the polygonal loop is a valid example. Furthermore, the Functional Fourier transform to the momentum loops would require path integration over coordinate loops $C$, which would make the relevant loops Brownian paths rather than smooth loops. The existence of the dot functional derivatives $\ff{\dot C(\theta)}$ will be required, but these derivatives exist for finite vorticity regardless of the smoothness of the loop itself. This is based on a Stokes theorem, valid for any continuous loop bounding a surface.
\end{remark}
\subsection*{\YM{} gradient flow}
Here is the definition of the  \YM{} gradient  flow
\begin{align}
   & \partial_\tau A_\nu = \alpha [D_\mu, F_{\mu \nu}],\\
   & F_{\mu \nu} = [D_\mu, D_{\nu}];\\
   & D_\mu = \pd{\mu} + A_\mu;
\end{align}
Here, the gauge field $A_\mu$ belongs to some Lie algebra, $ A_\mu = \I \sum T_a A^a_\mu$, which we do not need to specify.
The equation for the trace  of the ordered path integral (with $\mathbb P$ denoting a standard path ordering operator in QFT textbooks)
\begin{equation}
    W[C] = \tr \mathbb P \exp{\int_C d x_\alpha  A_\alpha}
\end{equation}
has a similar geometric form
\begin{align}
    \pd{\tau}W[C] = \alpha \tr \mathbb P \int_C d x_\alpha \left[D_\beta ,[D_{\beta}, D_{\alpha}]\right] \exp{\int_C d x_\alpha  A_\alpha}
\end{align}
This formula does not represent a closed equation for the loop functional, as the right-hand side still depends on the dynamic variables in coordinate space, satisfying the nonlinear partial differential equations we intended to solve in the first place.

\subsection*{Operator identity transforms nonlinear flow into  loop space diffusion}
The transformation of both equations to the loop space diffusion equation
is based on the operator calculus invented by Feynman \cite{Feynman1951}.
As applied to our case of the loop functionals for fluid dynamics
and the gauge theory, it allows us to write both of them as the
path ordered exponential times the unit operator
\begin{align}
    & \tilde \Psi[C] \otimes \mathbb I = \mathbb P \exp{\oint d x_\alpha D_\alpha(x_0)};\\
    & W[C] \otimes \mathbb I = \tr \mathbb P \exp{  \oint d x_\alpha D_\alpha(x_0)};
\end{align}
This fundamental geometric identity, proven in the \cite{migdal2025SQYMflow}
and our Appendix A, is identical in abelian and non-abelian cases, except
that a group space trace is taken in the \YM{} case. In both cases, it reveals the
geometric meaning of the loop functional as a parallel transport of the
\textbf{covariant derivative operator} around a loop $C$, reducing to a
c-number for a closed loop. The reduction of the ordered exponential to
a c-number is also supported by the Magnus expansion \cite{Magnus1954}.
\begin{remark}
The term c-number in this context means any operator commuting with all components of covariant derivative operator $D_\alpha(x_0)$, in particular, it could be an arbitrary complex number.
   This distinction between conventional definition of parallel transport and this operator definition is subtle but critical. 
   In much of the mathematical
literature, such as \cite{Gockeler1987}, the Wilson loop (the L.H.S. of our identity) is taken as
the \textit{definition} of parallel transport. Our proof, however,
establishes this correspondence as a rigorous \textit{equality} derived
from the more fundamental, anchored operator (the R.H.S.). This
approach mirrors a historical pattern in theoretical physics where a
concept, once viewed as a powerful analogy, is later shown to be a
provable, first-principles identity. 
\end{remark}

Now, the covariant derivative operator $D_\mu$ in both theories can be replaced by a functional derivative acting on the ordered exponential involving that same operator (remember that the operators $D_\mu(x_0)$ in this identity refer to the (arbitrary) origin $x_0 = C(\theta=0)$ on the loop.
In this operator form, the functional derivatives $\ff{\dot C_\mu(t\pm)}$ bring down the covariant derivative operator before or after the ordered product, which makes no difference due to the cyclic symmetry of the trace:
\begin{align}\hypertarget{dotDerDmu}{}
\label{dotDerDmu}
   & \ff{\dot C_\mu(t\pm)}W(C(.),\tau) =\VEV{\frac{1}{N}\tr D_\mu(C(t\pm)) \right.\nonumber\\
   &\left. \mathbb P\exp{\int_t^{t+2\pi} d s \dot C_\mu(s) D_\mu(x_0)}}
\end{align}
In this formula, we used cyclic symmetry of the trace and invariance of the Wilson loop with respect to the choice of the origin on a circle $(0, 2 \pi) \Ra (t, t + 2 \pi)$.  
We call such functional derivatives \textbf{the dot derivatives}. 

Applying the same formula for the dot derivative three times and properly contracting tensor indices, we bring the triple commutator $\left[D_\beta ,[D_{\beta}, D_{\gamma}]\right]$ from the exponential, which results in the diffusion equation  for $\tilde\Psi[C]$:
\begin{equation}\label{LoopEq}
\partial_t \tilde\Psi[C, t] = \nu \mathcal{L}_C \tilde\Psi[C, t]
\end{equation}
with the diffusion operator
\begin{align}\hypertarget{diffusionOperator}{}
\label{diffusionOperator}
   & \mathcal L_C = \oint d \theta \dot C_\nu(\theta) \hat L_\nu(\theta);\\
   &\hat L_\nu(\theta) = T^{\alpha\beta\gamma}_\nu \frac{\delta^3}{\delta \dot C_\alpha(\theta-0)\delta \dot C_\beta(\theta)\delta \dot C_\gamma( \theta+0)};\\
   &T^{\alpha\beta\gamma}_\nu =\delta_{\alpha\beta}\delta_{\gamma\nu}+ \delta_{\gamma\beta}\delta_{\alpha\nu}-2\delta_{\alpha\gamma}\delta_{\beta\nu};
\end{align}
All three arguments of the third functional derivative tend to $\theta$ in the specified order. The dot derivatives generate covariant derivatives inside the ordered product in the Wilson loop, and contraction with the tensor $T$ arranges these operators into a triple commutator 
\begin{align}
    T^{\alpha\beta\gamma}_\nu D_\alpha D_\beta D_\gamma =\left[D_\mu,\left[D_\mu,D_\nu\right]\right];
\end{align}
\begin{remark}
    The notion of the diffusion equation may be confusing, as until this moment we were studying a simple phase factor (exponential of circulation for a given velocity field around a fixed or moving loop).  The diffusion implies a statistical process with some probability distribution of random variables. How did we switch from a particular solution of the \NS{} (or \YM{}) equations to the diffusion process.? Where is the origin of randomness here? This fundamental approach was pioneered by Eberhard Hopf in early fifties \cite{Hopf1952}, and it assumes that initial data for velocity field was a statistical distribution rather than a smooth initial field assumed by naive mathematicians as a Cauchy data. In the physical world, water only flows above zero degrees Celsius (ice does not flow, let alone turbulence). Finite temperature means Gibbs statistical ensemble for initial velocity field. These thermal fluctuations, though smaller than the turbulent ones, can trigger the spontaneous stochasticity of the flow \cite{Bernard1998JStatPhys,Thalabard2020CommPhys}. The argumentation of Hopf was more abstract, but it was the same idea  of statistical distribution of velocity fields, inherited from initial conditions and evolving by the \NS{} equations. Furthermore, he expressed the idea of a turbulent attractor, which was a prophetic vision of modern theory of turbulence, with the Euler ensemble being precisely the loop scape implementation of Hopf's turbulent attractor.  Technically, the loop equation being a \emph{linear} functional differential equation, obeys the superposition principle. Therefore, being derived for an \emph{arbitrary solution} of the \NS{} equation, it also applies to any linear superposition, in particular, to the statistical average of the loop functional over initial Gibbs distribution $\exp{-\beta\int v^2}$ at $t=0$.
\end{remark}
\subsection*{Diffusion or quantum mechanics?}

The loop equation is formally real, yet the loop operator $\mathcal{L}_C$ is a third-order \emph{anti-Hermitian operator}. Consequently, none of the standard properties of elliptic operators associated with conventional diffusion apply to this loop space dynamics.

In fact, this system is more accurately interpreted as quantum mechanics governed by a Schrödinger equation:
\begin{align}\label{LoopEq}
&\I \nu \partial_t \tilde\Psi[C, t] =  \mathcal{H}_C \tilde\Psi[C, t];\\
& \mathcal{H}_C = \I \nu^2 \mathcal L_C
\end{align}
The viscosity $\nu$ possesses the same physical dimensions $[L]^2/[T]$ as the kinematic action (analogous to Planck's constant $\hbar$ per unit mass), and it plays a strictly analogous role in this loop space dynamics. The Hamiltonian is linear in the canonical coordinates $C(\theta)$ and cubic in the canonical momenta $\delta/\delta C(\theta)$, scaling as $C^{-2}$. This resembles the kinetic energy of a free particle, but without the requirement of positivity.

There are zero modes, which are responsible for the fixed points in the Yang-Mills case. However, in the Navier-Stokes case, there is only a trivial zero mode $\Psi[C]= 1$.

Crucially, the wave function is complex and exhibits oscillations, a hallmark of quantum mechanics. These oscillations manifest in observable correlation functions (behind the dominant power decay asymptotic terms) as genuine interference effects within this loop space quantum mechanics. \textbf{It is important to clarify that these quantum features are intrinsic to the continuum loop dynamics and do not arise from the discrete number-theoretic structure of the regular star polygons.}

The turbulent limit $\nu \to 0$ corresponds to the WKB limit of this loop space quantum mechanics. This analogy is the key to the analytic solution of the turbulence problem described in this review.

In the remainder of this paper, we shall refer to this dynamics as "loop space diffusion" to maintain familiarity for practitioners of fluid dynamics. However, make no mistake: this is an \emph{imaginary} diffusion, better known as quantum mechanics.

This raises an intriguing philosophical question: how can a deterministic classical system exhibit quantum interference effects?
First, the system is not truly deterministic in the reversible sense: the imaginary part of the loop functional $\Psi[C]$ is directly related to energy dissipation, leading to a violation of time-reversal symmetry. The probabilistic distribution of dissipative turbulent flow possesses all the features of a quantum mechanical wave function, including the quantum interference of amplitudes in the sum over histories. The Schrödinger equation here is not a poetic analogy but an exact mathematical equivalence, much like the thermal radiation of black holes.

Second, while complex numbers may appear to be a mathematical artifice, they are as fundamental here as the complex amplitudes and phases of electromagnetic waves in \emph{classical electrodynamics}. The phenomenon of wave interference is an objective physical process, and complex amplitudes are the necessary mathematical tools to describe it. The same logic applies to the circulation dynamics in turbulent flow. The interference is real; its mathematical mechanism is identical to that of quantum mechanics. Whether one labels it "quantum" or "dissipative statistical" is merely a matter of semantics.

\subsection*{A Note on Dot Derivatives and Kinematical Discontinuities}

It is important to clarify the nature of the "dot derivatives" ($\delta/\delta\dot{C}(\theta)$) that form the basis of loop space calculus. These are the standard functional derivatives with respect to the velocity of a periodic trajectory, familiar from the variational principles of classical mechanics (e.g., the Euler-Lagrange equations).

The profound subtlety arises when multiple derivatives are taken at the same parameter point $\theta$ along the loop. The result depends on the order in which the limit of coincident points is taken. For instance, the action of a second derivative depends on whether we take the points as $(\theta, \theta+\epsilon)$ or $(\theta+\epsilon, \theta)$ in the limit $\epsilon \to 0$.

This ordering dependence results in a finite discontinuity. This discontinuity is not a pathology or a short-distance singularity of the underlying field that needs to be regularized. Rather, it is a fundamental and universal \textit{kinematical property} of loop functionals.

The difference between the two possible orderings of a second dot derivative is directly proportional to the commutator of the underlying covariant derivatives. This mathematical structure precisely encodes the essential physics:
\begin{itemize}
    \item In \textbf{fluid dynamics}, this commutator is the \textbf{vorticity tensor}, $\omega_{\mu\nu}$.
    \item In \textbf{Yang-Mills theory}, it is the \textbf{field strength tensor}, $F_{\mu\nu}$.
\end{itemize}
Thus, the non-commutativity of the dot derivatives at a single point is the loop-space representation of the local field strength. This effect would be present even for the simplest non-trivial fields, such as a constant field strength in Yang-Mills theory or a rigid body rotation ($\vect{v} = \frac{1}{2} \boldsymbol{\omega} \times \vect{r}$) in hydrodynamics, where the fields themselves are perfectly smooth. These discontinuities are the mechanism by which the local, rotational dynamics of the field are captured by the calculus on the one-dimensional loop.

Another subtlety is the operator nature of these dot derivatives. Every functional derivative $\delta/\delta\dot{C}(\theta)$ brings down from the path-ordered exponential a covariant derivative operator. The resulting object is no longer a c-number, as this covariant derivative acts on all the factors to the right in the path-ordered exponential. After commuting with all these factors in the ordered product $(1 + d\theta \dot C_\mu(\theta) D_\mu(x_0))$, this covariant derivative operator moves to the end of the path ordered product and "hangs" there, making the whole expression an operator in Hilbert space rather than a c-number.

The c-number is restored after taking the discontinuity of the first dot derivative or the antisymmetrization by tensor indices of the second dot derivative. After this antisymmetrization, the product of two operators $D_\mu D_\nu$ in front of the path-ordered exponential becomes a c-number commutator $[D_\mu,D_\nu]$, and it no longer differentiates the path-ordered exponential to its right. Now we get back the original Wilson loop and its dot derivatives. 

So, we hop into a Hilbert space at each dot derivative but come back to a world of numbers after taking discontinuity or antisymmetrization. This transformation also applies to the triple commutator: it is a c-number only after the symmetrization over tensor indices by multiplication by the tensor $T^{\alpha\beta\gamma}_\nu $.
\subsection*{The universal loop equation with varying initial data}
This equation governs the evolution of the system's complete statistical state.

With finite viscosity, there is a universal diffusion equation in loop space. The turbulent limit $\nu \to 0$  corresponds to the WKB limit in that diffusion equation, and it is determined by the zero modes of the operator $\mathcal{L}_C$.

In the case of gauge theory, the \emph{same} equation describes the evolution of the Wilson loop in \YM{} gradient flow. The difference lies in the initial data.
The general solution of the loop equation in both cases can be written as
\begin{align}
   & \Psi[C] = \exp{\nu t \mathcal{L}_C } \Psi_0[C];\\
   & W[C] = \exp{\alpha t \mathcal{L}_C } W_0[C]
\end{align}
The initial data are, of course, different, corresponding to the distributions of the two different vector fields.
\begin{align}
   & \Psi_0[C] = \VEV{\exp{\frac{\I \oint_C d x_\mu v^{0}_\mu(x) }{\nu}}}_{v^0};\\
    & W_0[C] = \VEV{\tr \mathbb P \exp{ \oint_C d x_\mu A^{0}_\mu(x) }}_{A^0}
\end{align}
The mathematical tool that enables this formulation is a \textbf{loop space calculus}. Unlike previous approaches that required adding singular cusps to loops, our calculus operates entirely within the manifold of smooth loops. Variational operators are defined via derivatives with respect to the loop's velocity profile, $\dot{C}(\theta)$. This method is free of the singularities and ambiguities that plagued earlier attempts, yielding a well-defined operator $\mathcal{L}_C$. The technical details of this calculus are provided in Appendix A.

The most important aspect of this new loop equation is that it is analytically solvable by functional Fourier transform in loop space.

\section{Momentum loop equation}
\pdfbookmark[1]{Momentum loop equation and decaying gradient flow}{Anzatz}

Now let us return to the nonsingular loop equation \eqref{LoopEq}.
The diffusion equation in any linear space can be exactly solved by the Fourier transform, and the loop space is no exception. However, its infinite dimensionality makes the solution quite nontrivial.

\hypertarget{sub-the-plane-wave-in-loop-space}{}
Let us  write down an  Ansatz for a solution in the form of the momentum loop equation (MLE)
\begin{align}\hypertarget{Anzatz}{}
\label{Anzatz}
    & W[C,\tau] = \VEV{\exp{\I \int_0^{2\pi}  d \theta \dot C_\mu(\theta) P_\mu(\theta,\tau)}}_{P(\tau)};
\end{align}
Substituting this Anzatz into the loop equation \eqref{LoopEq}, we find that this equation requires the following evolution of momentum loop
\begin{align}
    \hypertarget{MLE}{}
\label{MLE}
    & \frac{\pd{\tau} \bar P}{\nu} =  (\Delta P\cdot \bar P) \Delta P - (\Delta P)^2 \bar P;\\
    & \bar P =  \frac{P(\theta+)+P(\theta-)}{2}; \\
    & \Delta P =  P(\theta+)-P(\theta-); 
\end{align}
The brackets $\VEV{}_{P(\tau)}$ correspond to the averaging over an ensemble of solutions of the time evolution of $P_\mu(.,\tau)$ described by the above equation.
\begin{remark}
    Let us clarify this important point. This equation for $P(\theta,\tau)$ is an ordinary differential equation with time derivative $\pd{\tau}  P(\theta,\tau)$ being a nonlinear functional of $P$  depending on infinitesimal neighbors on a unit circle  $P(\theta\pm 0,\tau)$. This equation is singular but deterministic. The randomness comes from the initial data $P(\theta,0)$, which are random variables with the distribution given by Fourier transform of the initial $\Psi(C)$. This distribution was computed in \cite{migdal2025duality}, but we do not need it here. The idea behind this MLE is that this solution $P(\theta, \tau)$ starts from a specific random data $P(\theta, 0)$, determined by transforming the initial Gibbs average $\VEV{\Psi(C)}_{\mathrm{Gibbs}}$ into momentum loop space, but in course of evolution, it is attracted by a fixed point (the Hopf's turbulent attractor). This fixed point is a \emph{degenerate} solution of the MLE, depending upon infinite number of arbitrary parameters. In other words, this fixed point is a set or a manifold, just like the Gibbs distribution was a manifold (the sphere of constant energy in the infinite dimensional space of the velocity field). Where is this attractor located? Not in the physical space, like inside the bulk of the flow. Not even in the loop space $C(\theta)$-- it is located in the momentum loop space $P(\theta)$. The momenta are related to coordinates as hamiltonian conjugates, which means that discrete localized structures in momentum space correspond to continuous distributions in coordinate space, -- here in the loop space. So, the regular star polygons of the Euler ensemble sit in abstract space, not reflecting the discrete structures in physical space (with a notable exception of the passive scalar we consider later in this review).
    Given the attractor manifold, we follow ergodic theory by assuming that the time evolution uniformly covers this manifold with an invariant measure determined by its internal mathematical symmetries. The miracle is that in the \NS{} MLE, this manifold can be analytically found, its invariant measure computed using modern Number Theory, and the emerging turbulent distribution investigated in the local limit, when it also has infinitely many degrees of freedom. The idea of a fixed point is a basis of Wilson's RG approach to critical phenomena, where it is more a philosophy than an analytical method. Not so in the turbulence problem! Here we have a priceless gift of analytic solution rooted in the Number Theory.
\end{remark}

This Anzatz \eqref{Anzatz} is the loop space version of the plane wave. The loop equation \eqref{LoopEq} involves only dot functional derivatives; therefore, an Anzatz would exactly satisfy the loop equation, with each dot derivative $\ff{\dot C_\nu(\theta)}$ equivalent to multiplication of $\I P(\theta)$.
There is a one-to-one correspondence between the original \NS{} equation and this algebraic relation between momentum loops and their discontinuities.
The nonlinearity of the \NS{} equation results in nonlinearity in the momentum loop equation, with the different terms in the \NS{} equation combined into a simple cubic polynomial, reflecting the structure of the triple commutator of covariant derivatives.

The relation between the loop operator and the discontinuity of the momentum loop $\Delta P_\nu(\theta)$ was discovered and investigated in earlier papers in QCD \cite{MLDMig86, SecQuanM95, Mig98Hidden}. The Momentum loop equation for QCD involves contact terms that provide boundary conditions in loop space for self-intersecting loops. We are not studying these terms in the present paper, restricting ourselves to the \YM{} gradient flow. For QCD, this equation corresponds to the WKB limit of large loops without self-intersections, when some form of the area law is expected as an asymptotic solution.

The mathematical meaning of these discontinuities is well known.
\section{Application I: Decaying Hydrodynamic Turbulence}

We first apply the framework to decaying homogeneous isotropic turbulence—Feynman's "oldest unsolved problem." The momentum loop equation \eqref{MLE} admits a universal, long-time attractor solution that is independent of initial conditions. The time dependence immediately follows from the fact that the right side of \eqref{MLE} is a cubic homogeneous functional of $P$, and the left side is the time derivative of $P$. The equation then becomes a nonlinear relation for the coefficient in front of the power factor
\begin{align}\label{Psol}
   & P_\mu(\theta,\tau) = \frac{f_\mu(\theta)}{\sqrt{2\nu (\tau + \tau_0)}};\\
   \label{feq}
   & \bar f \left(\Delta f^2-1\right) = (\Delta f  \cdot \bar  f )  \Delta f
\end{align}
For the loop functional, it reduces to the following.
\begin{align}\label{Psisol}
   \tilde \Psi[C] = \VEV{\exp{ \frac{\I }{\sqrt{2\nu (\tau + \tau_0)}}\sum_k  f_\mu(\theta_k) \Delta C_\mu(\theta_k)}}_{\mathcal E}
\end{align}
where the averaging is over the solutions of the recurrent equation \eqref{feq}. The equally spaced points on a unit circle $\theta_k = \frac{2 \pi k}{N}$ approximate the infinitesimal step $\theta\pm 0$ involved in the original loop equation. Such a function $f(\theta)$, with discontinuity at every angle, can exist only in the sense of distributions, which is why this limiting procedure is required. These subtle issues of discretization of the loop equation are discussed in the recent paper  \cite{DeLellisInprep}, where they are justified under certain assumptions about the continuity of the loop and the flow.

The limit $N \to \infty$ should be taken in the end \emph{at fixed turbulent viscosity} $\tilde \nu = \nu N^2$. The last relation was derived in \cite{migdal2023exact} and used later in \cite{migdal2024quantum}. It guarantees a finite limit for observable correlation functions, which was analytically computed in these papers, using QFT methods combined with number theory.
\begin{remark}
    The requirement of fixed turbulent viscosity arises from the necessity that the energy dissipation rate in the infinite system
    \begin{align}
        \mathcal{E} = \frac{\nu \int_V \omega^2}{V}
    \end{align}
    stays finite in the turbulent limit $\nu \to 0$, taken simultaneously with the local limit $N \to \infty$ when the approximating polygon becomes a continuous loop $C$.  In this limit, our discrete MLE tends to the solution of the continuum loop equation. The scaling relation $ \nu \propto 1/N^2$ follows from the estimate \cite{migdal2023exact, migdal2024quantum} using the second area derivative to compute $\omega^2$ and shrinking the loop to a point. In this limit, the statistical average of  \eqref{Anzatz} is determined by the WKB asymptotics, and this leads to dramatic simplification, eventually producing exact finite result for the vorticity correlator. This result involves the average over an ensemble of $\mathbb{Q}$ rational numbers, eventually leading to beautiful Mellin-Barnes integrals involving ratios of Riemann zeta functions \cite{migdal2024quantum},section IX.
\end{remark}
\subsection*{Recurrent equation on a circle}
The last equation \eqref{feq} for $f_\mu(\theta)$ can be solved exactly, by a certain limiting procedure.
First, we observe that the left and right sides are vector products with some scalar coefficients. Unless both of these coefficients vanish,
 these vectors are collinear.
But in that case, the area derivative of the Wilson loop identically vanishes at every point on an arbitrary loop $C$. Such would be a trivial solution in which the gauge field is a pure gauge (a velocity field is purely potential, without any vorticity that could lead to turbulence).

The nontrivial solutions, with finite vorticity, all correspond to both scalar coefficients vanishing for every angle $\theta$. We can rewrite these equations as
\begin{align}\label{recEq}
     & (f(\theta+) + f(\theta-))\cdot(f(\theta+) - f(\theta-))=0;\\
     & (f(\theta+) - f(\theta-))^2 =1;
\end{align}
The first equation can be rewritten as the continuity of the length of the vector $f(\theta)$
\begin{align}
     f(\theta+)^2 = f(\theta-)^2
\end{align}
We conclude that these vectors are located on a sphere:
\begin{align}\hypertarget{fsol}{}
\label{fsol}
     f(\theta) = R n(\theta); \textit{ where } n(\theta) \in \mathbb S_2
\end{align}
The second equation relates the radius of the sphere to the constant angle between the consecutive vectors
\begin{align}
     1-n(\theta-)\cdot n(\theta+)= \frac{1}{2 R^2 } = \const{}
\end{align}
There is also an important requirement of periodicity
\begin{align}\hypertarget{periodicity}{}
\label{periodicity}
     n(\theta+ 2 \pi) = n(\theta)
\end{align}
This solution in three dimensions is the \textbf{Euler ensemble}.

\subsection*{The Euler Ensemble: A Universal Turbulent Attractor}
In this solution, $\vect{f}(\theta)$ is a universal fractal curve constructed as the continuum limit ($N \to \infty$) of a random walk on a regular star polygon $\{q/p\}$. Its vertices are given by:
\begin{align}
    &\vect{f}\left(\frac{2 \pi k}{N}\right) = \hat{\Omega} \cdot\{R\cos(\alpha_k), R\sin(\alpha_k), 0\};\\
    & R = \frac{1}{2 \sin(\beta/2)};
    \label{EulerEnsemble}
\end{align}
where the rotation matrix $\hat{\Omega} \in SO(3)$, the angle step $\beta = 2\pi p/q$ with $p/q \in \mathbb{Q}$, and the cumulative angle $\alpha_k = \beta \sum_{l=1}^k \sigma_l$ with $\sigma_l = \pm 1$ define the random walk.

The recurrent equations \eqref{recEq} are satisfied in an obvious geometric way: with this radius $R $ the sides have unit length: $$R \abs{e^{\I \alpha_{k+1}} - e^{\I \alpha_k}} =R \abs{e^{\I\beta \sigma_k} -1} =1.$$
The Euler ensemble is not a single configuration but a degenerate manifold of solutions, a "fixed stochastic trajectory."
The statistical state of turbulence is a distribution over all choices of $(\hat{\Omega}, p, q, \{\sigma_k\})$.

The periodicity condition $\alpha_N-\alpha_0 = \beta \sum\sigma_k = 2 \pi r$ imposes a number-theoretic constraint on the geometry of the walk: for a walk to close after traversing N steps, the rotation angle $\beta$ must be a rational multiple of $2\pi$, i.e., $\beta=2\pi p/q$. The requirement that the polygon $\{q/p\}$ be irreducible (i.e., not a simpler polygon traced multiple times) restricts $p$ and $q$ to be coprime, meaning the number of unique star polygons for a given denominator $q$ is precisely $\varphi(q)$, the Euler totient function.

This point is crucial-- the origin of the link between turbulence and number theory. We are limited to the manifold $\mathbb Q$ of rational numbers, which eventually leads to all the "quantum" effects in the solution of decaying turbulence.
The statistics of turbulence are thus described by a uniform distribution over all such polygons and all possible random walks on them. This discrete, number-theoretic structure is the hallmark of spontaneous quantization.

The distribution of the variable $X(p,q) =\cot(\pi p/q)^2/N^2 = \left( 4 R^2 -1\right)/N^2$ for co-prime $1 \le p<q < N$ was computed in \cite{migdal2024quantum} by advanced number theory methods. In the statistical limit $ N \to \infty$, this distribution remains discontinuous
 \begin{align}\label{rhodist}
    & f(X)= \left(1-\frac{\pi ^2}{675 \zeta (5)}\right)\delta(X) +\frac{\pi^3}{3} X\sqrt{X}\Phi\left(\floor*{\frac{1}{\pi \sqrt{X}}}\right);
\end{align}
with $\Phi$ the totient summatory function:
\begin{align}
     \label{PhiDef}
    & \Phi(q) = \sum_{n=1}^q \varphi(n) ;\\
    & \varphi(m) = m \prod_{p|m}\left(1 - \frac{1}{p}\right)
\end{align}
The discontinuities of $f(X)$ at quantized values are related to famous Euler totients $\varphi(n)$
\begin{align}
    & \Delta f(X_n) = \frac{\varphi(n)}{3 n^3}  ;\\
    & X_n = \frac{1}{\pi^2 n^2}
\end{align}
This distribution has finite support $ 0 < X < \frac{1}{\pi^2}$ and condenses to a power law as $ f(X) \propto \sqrt{X} $  at  $X \to 0$.

The Euler ensemble is remarkably universal. As we show in the Appendix, it also describes the solution of the loop equation for the static loop (not moving with the flow). Explicit computation shows that the advection term $v \omega$ present in the static loop equation, when integrated over the loop using the solution for vorticity and (by inverting the curl operator) for velocity for the Euler ensemble, becomes a total derivative of a periodic function, so that the integral over $\theta$ yields zero. This nontrivial cancellation was not built into the solution for the liquid loop but emerged as a free bonus, suggesting a deeper reason for this solution -- not yet completely understood.
\subsection*{Historical note (Wikisource)}
The study of star polygons dates to antiquity with Pythagoras and his school. The subject was later notably taken up by Thomas Bradwardine, whose 14th-century mathematical work on these figures, conducted while he was Archbishop of Canterbury, is a testament to the era's polymathic tradition.
The statistical limit of random walk on regular star polygons has never been solved before, to the best of our knowledge.
\begin{figure}[h!]
     \centering
     \includegraphics[width=0.8\columnwidth]{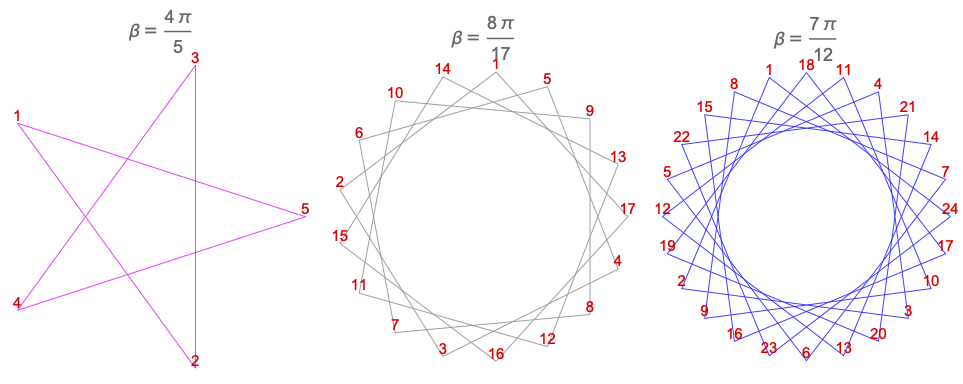}
     \caption{Examples of the regular star polygons, $\{q/p\}$, that form the discrete target space of the dual string theory. The turbulent state, or Euler ensemble, is a statistical average over random walks on all such polygons.}
     \label{fig:RegularPolygons}
\end{figure}
\begin{figure}[h!]
\centering

\includegraphics[width=\linewidth,height=0.5\textheight,keepaspectratio]{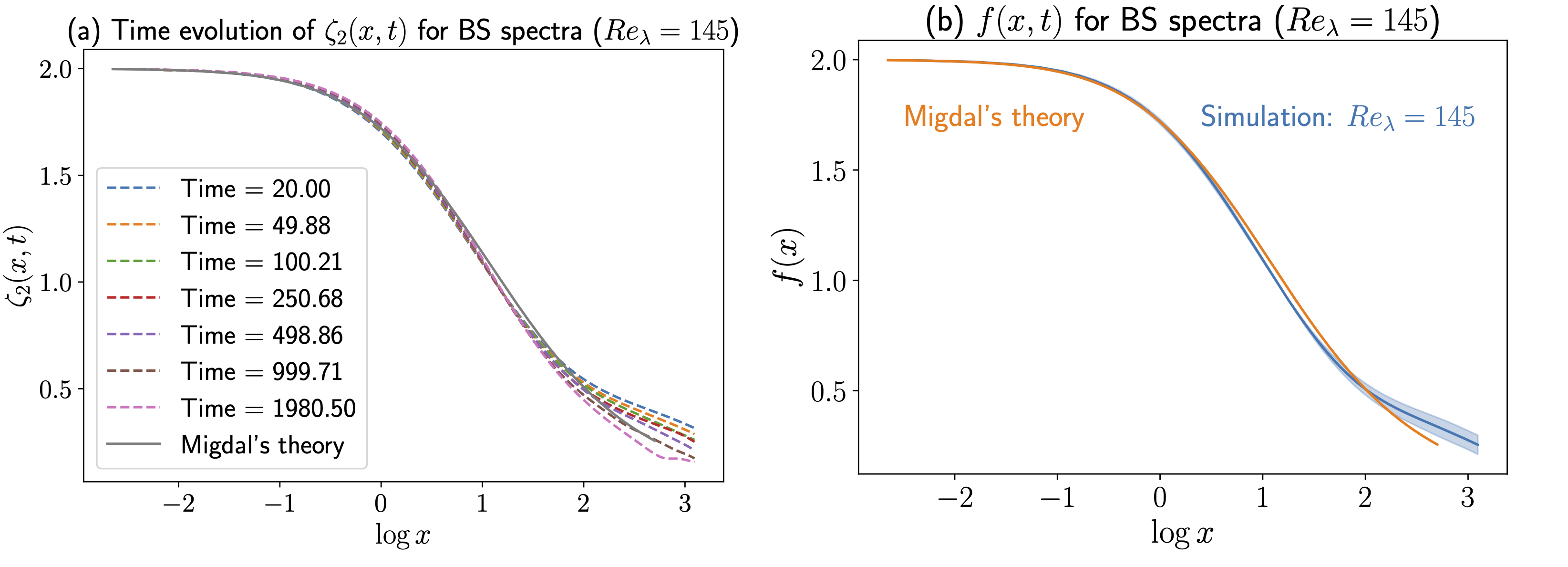}

\includegraphics[width=\linewidth,height=0.5\textheight,keepaspectratio]{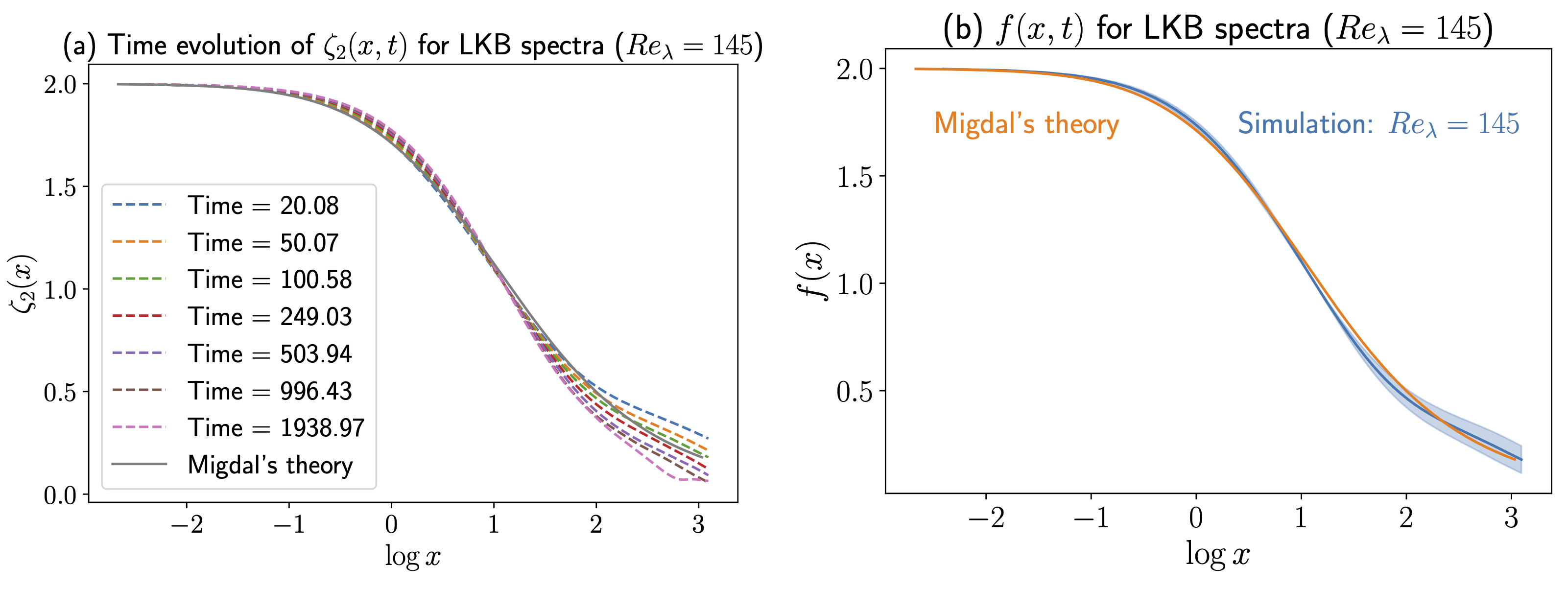}

\vspace{4mm}

\includegraphics[width=\linewidth,height=0.3\textheight,keepaspectratio]{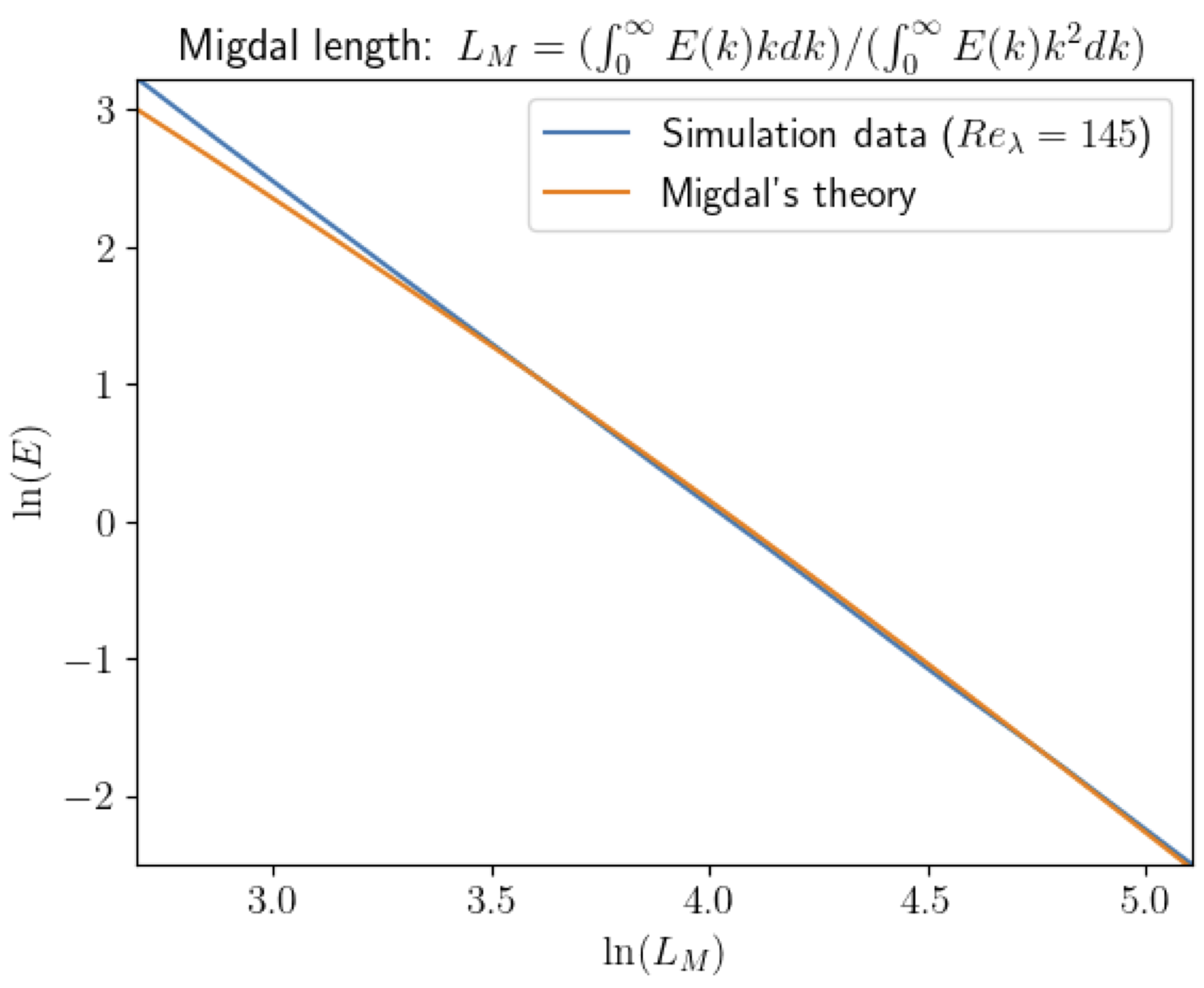}

\caption{Comparison of the Euler-ensemble prediction with the \(4096^3\) DNS of Sreenivasan and Rodhiya \cite{SreeniAkash2025}. \textbf{Top:} Effective index \(f(x)=\langle r\partial_r\log(\Delta v^2)\rangle\) for the Birkhoff--Saffman (\(k^2\)) initial spectrum. \textbf{Middle:} The same quantity for the Loitsyansky--Kolmogorov--Batchelor (\(k^4\)) initial spectrum. In both cases the theory reproduces the observed universal curvature. \textbf{Bottom:} BS total-energy decay at \(Re_\lambda=145\). The DNS deviates from the leading asymptotic term \(E\propto L_M^{-5/2}\) but agrees well with the full theoretical curve including subleading Mellin--Barnes corrections.}
\label{fig:DNSFit}
\end{figure}
\subsection{Key Prediction 1: The Energy Spectrum and Intermittency}
The Euler ensemble provides a complete, parameter-free prediction for all statistical properties of decaying turbulence. Most notably, it determines the velocity correlation functions and the energy spectrum \cite{migdal2024quantum}. The computations are rather heavy; in addition to above distribution of the radius $R$ of the regular star polygons, they involve some QFT technology, such as path integrals in the WKB limit, related to the classical trajectory for the Ising spin density $\alpha(k/N)$ and the functional determinant for harmonic fluctuations around this trajectory, computed by the zeta regularization method.

The results for vorticity correlation functions in Fourier space are Mellin-Barnes integrals, with a meromorphic function involving the exponential of another integral arising from the zeta-regularized log-determinant of the aforementioned quadratic form of harmonic fluctuations around the classical path. These integrals, though, are calculable with high precision by \Mathematica{}.

These two key observables—the \textbf{energy spectrum} in Fourier space and the \textbf{velocity moments} in coordinate space—are not independent. They are two different representations of the same underlying velocity correlation function. The energy spectrum $E(k,t)$ is directly related to the Fourier transform of this function, while the second-order velocity moment $\langle (\Delta \vect{v})^2 \rangle(r,t)$ is its manifestation in real space. Therefore, the theoretical prediction for the scaling exponents is unified, governing both the shape of the energy spectrum in $k$-space and the scaling of the velocity moments in $r$-space.

The second moment of the velocity difference, for instance, is given by an infinite series of power laws, determined by a Mellin-Barnes integral:
\begin{equation}
\langle (\Delta \vect{v})^2 \rangle(r,t) \propto \frac{1}{t} \sum_{\text{poles } p_i} \text{Res}[V(p_i)] \left(\frac{r}{\sqrt{t}}\right)^{p_i}
\end{equation}
The function $V(p)$, derived from the theory (see Appendix C), has poles that define the universal intermittency exponents, $p_i$. This spectrum of indices is remarkable: it contains not only rational numbers but also an infinite series of complex-conjugate pairs, $p_n = 7 \pm i t_n$, where $\frac{1}{2} \pm i t_n$ are the \textbf{nontrivial zeros of the Riemann zeta function}. This prediction \cite{migdal2024quantum}, which is in excellent agreement with high-resolution direct numerical simulations (DNS), establishes a deep and unexpected connection between turbulence and number theory.

A direct test of this predicted functional form is the effective index, 
$f(x) = \langle r \partial_r \log(\Delta v^2) \rangle$, where $x = r/L(t)$. 
This test is shown in \textbf{Figure 2 (Top)}, which compares the 
parameter-free theoretical curve (orange line) to the latest $4096^3$ DNS data  \cite{SreeniAkash2025}. The agreement is excellent 
across the primary range. 

Deviations are visible at \textbf{large $\log(x)$}, which are expected. 
This right-hand part of the plot, where $x = r/L(t)$ is large, is 
contaminated by two distinct systematic effects. First, at any given time, 
the scales $r$ are approaching the finite simulation box size, leading to 
\textbf{lattice artifacts}. Second, the time-averaging includes data from 
early simulation times, when $L(t)$ is still small and influenced by the 
\textbf{initial conditions}; this small $L(t)$ also shifts the data to large $x$.

The true statistical errors are understood to be much smaller than the 
displayed error bands, which are inflated by these systematic, 
non-statistical artifacts. We display the full curve, including these 
deviations, to avoid any temptation to discard data that does not fit 
the theory. Instead, these deviations at large $x$ are themselves 
understood as known, physical consequences of the simulation's constraints. 
Even with these effects, the theoretical 
curve remains within approximately two standard errors of the data, 
indicating strong agreement.

Furthermore, the log-log slope of $E(t)$ versus $L(t)$ in the DNS data shows a
slight, systematic deviation from the \textbf{leading-order theoretical exponent}
of $-5/2$. This deviation is not a flaw; it is perfectly explained by the theory.

As shown in \textbf{Figure 2 (Bottom)}, the leading-order term (gray dash-dotted
line) acts as an asymptote, while the \textbf{complete theoretical curve}
(orange line)—which includes all sub-leading exponents from the numerically
evaluated Mellin-Barnes integral—provides a \textbf{precise match to the
$Re_{\lambda}=145$ DNS data} over the entire range. This fit, which matches
thousands of data points using only two physical scale parameters (for overall
energy and length), provides a high level of confidence in the solution.
It visually confirms that the observed energy decay law is composed of the
"ground state" term ($L_M^{-5/2}$) plus a hierarchy of calculable,
quantum-like corrections, just as the theory predicts.
\begin{remark}
    It may be worth noting that this superposition of leading plus subleading power terms $$ E(L) \to A L^{-5/2} + B L^{-11/2} + \dots$$ will not imitate a mixture of powers. Asymptotically, the effective index $$n(L) = LE'(L)/E(L) \to -\frac{5}{2} - \frac{3 B}{A L^3}+ \dots$$ will rather look like a leading index plus a decreasing powerlike correction, so this will not be a rotated line in a log-log scale but rather  a straight line with a leading slope, curved at the left end by a $1/L^3$ correction, just like it appears in Figure \ref{fig:DNSFit}.
\end{remark}
Presumably, the minor deviations from the theoretical curve at the far left
end ($ \log L_M < 3.2, \log E > 2$) are explained by the early stages of energy decay, when
statistical equilibrium has not yet been achieved. The initial K41 spectrum would
have contaminated the data, leading to deviations from the turbulent attractor.

\subsubsection*{LKB regime violation of universality as a boundary effect}
The Euler ensemble solution applies to decaying turbulence in an infinite system at an infinite Reynolds number. Still, as we have seen, the DNS for a finite grid with $ L=4096$ steps and Reynolds number $145$ already matches our prediction within statistical errors, up to the boundary corrections, showing that at large distances $r \sim L$ or small wavelengths $k \sim \pi/L$.

These boundary corrections are also related to the initial shape of the energy spectrum. The time decay of the energy spectrum starts at large $k$, where the K41 spectrum  $E(k,0) \sim k^{-5/3}$ was imposed as initial data for lack of a better  guess.

This spectrum decays at the high end, curves on a log scale, and effectively shifts all the points to the left. The lower part of the spectrum remains at its initial state at $t=0$. There are two popular choices:  Birkhoff-Saffman regime (BS): $E(k,0) \propto k^2$ and Loitsianskii-Kolmogorov-Batchelor (LKB) regime $E(k,0) \propto k^4$. The BS regime is the most general; it requires no parameter tuning, and it is apparently observed in physical experiments on decaying turbulence in water tanks, pipes, and wind tunnels. The experimental data, corrected for the boundary effects, perfectly match our predictions $E_{tot}(t)\propto t^{-5/4}$.

In the DNS of the $4096^3$ grid that we analyzed above, the BS regime also fits our theory very well. However, as observed in \cite{SreeniAkash2025}, with the LKB initial data, the fit is not as good; there are systematic deviations that the DNS authors fit to the LKB version of the K41 law. 
$$ E_{tot}(t) \approx t^{-10/7}.$$
On the other hand, the data for the second-order velocity moment $\langle (\Delta \vect{v})^2 \rangle(r,t)$ in the LKB regime \cite{SreeniAkash2025} matched our theory as well as they did in the BS regime.
Both observables being mathematical transformations of the same microscopic quantity—the decaying energy spectrum—this paradoxical discrepancy needs an explanation.

This explanation is simple: the effective index is the ratio of two integrals over the energy spectrum, with kernels supported on the bulk of the flow ($\pi/L \ll k \ll \pi/2$).  This region is supposed to be described by our turbulent attractor-- the Euler ensemble.

The total energy, as defined, equals the integral over the full spectrum. In the lattice system, this will be the sum for all quantized wavevectors $ k = \pi n/L$.  The first several terms in this sum are influenced by the initial $k^2$ or $k^4$ spectrum.

We do not expect our theory to match the low end of the energy spectrum: there are systematic errors of the lattice approximation to the true isotropic homogeneous turbulence. However, the theoretical spectrum grows at small $k$, so the total energy does not converge in the bulk. We had to cut off the spectral integral at some finite $k = k_0\sim 1/L$ in \cite{migdal2024quantum} to predict the scaling law $E_{tot}(t) \sim t^{-5/4}$. The coefficient in front of this power law is not universal: it is proportional to the corresponding power of $k_0$.

To avoid these apparent violations of universality due to contamination from the initial/boundary data, we switch to the enstrophy, which is proportional to the time derivative of energy, so it can serve equally well to describe the energy decay 
\begin{align}
& E'_{tot}(t) = -\nu En(t);\\
   & En(t) = \int_0^\infty d k k^2 E(k,t);
\end{align}
The scaling law for enstrophy in our theory is
\begin{align}
    En(t) \propto  (t+ t_0)^{-9/4} \propto L(t)^{-9/2}
\end{align}
The $10/7$ law  for the energy would translate into 
\begin{align}
   \text{K41: } En(t)\propto  (t+ t_0)^{-17/7} \propto L(t)^{-34/7}
\end{align}
The difference with the energy is that the enstrophy integral \emph{is dominated by  the bulk}, where the theory of turbulence would apply. Neither the boundary conditions nor the low end of the spectrum, contaminated by initial data, will contribute to the enstrophy integral. Moreover, the enstrophy, the mean square of the vorticity, does not depend on the potential part of the velocity field. 

It is this potential component that is most affected by the initial data for the BS or LKB regimes. The coefficient in front of $k^2$ is proportional to the square of mean velocity, and that in front of $k^4$ - to the square of mean angular momentum. In general, the $k^2$ term dominates, but if one tunes parameters so that there is no mean velocity, the $k^4$ term is left as the first term of expansion at small $k$. Obviously, neither of these terms has anything to do with the turbulent regime; these are just initial data for the potential part of the velocity field. These initial macroscopic motions are related to the dynamics of the vortex structures responsible for turbulence through the advection they induce. Thus, the solution for the velocity field will initially differ in these two regimes (BS and LKB), but we expect the vortex dynamics to eventually settle on the turbulent attractor described by the Euler ensemble.

The DNS data can be safely summed without filtering out the lattice artifacts at $k \sim 1/L$.  The integral will converge in the bulk, where the turbulence theory applies.
Thus, we can use the entire dataset in the turbulent range where statistical equilibrium is reached, but the turbulent kinetic energy has not yet dissipated. That will give us better statistics than for the bulk energy decay. The enstrophy test is therefore a direct probe of the turbulent attractor, uncontaminated by boundary or initial‑condition effects.

\begin{figure}[h!]
\includegraphics[width=0.9\columnwidth]{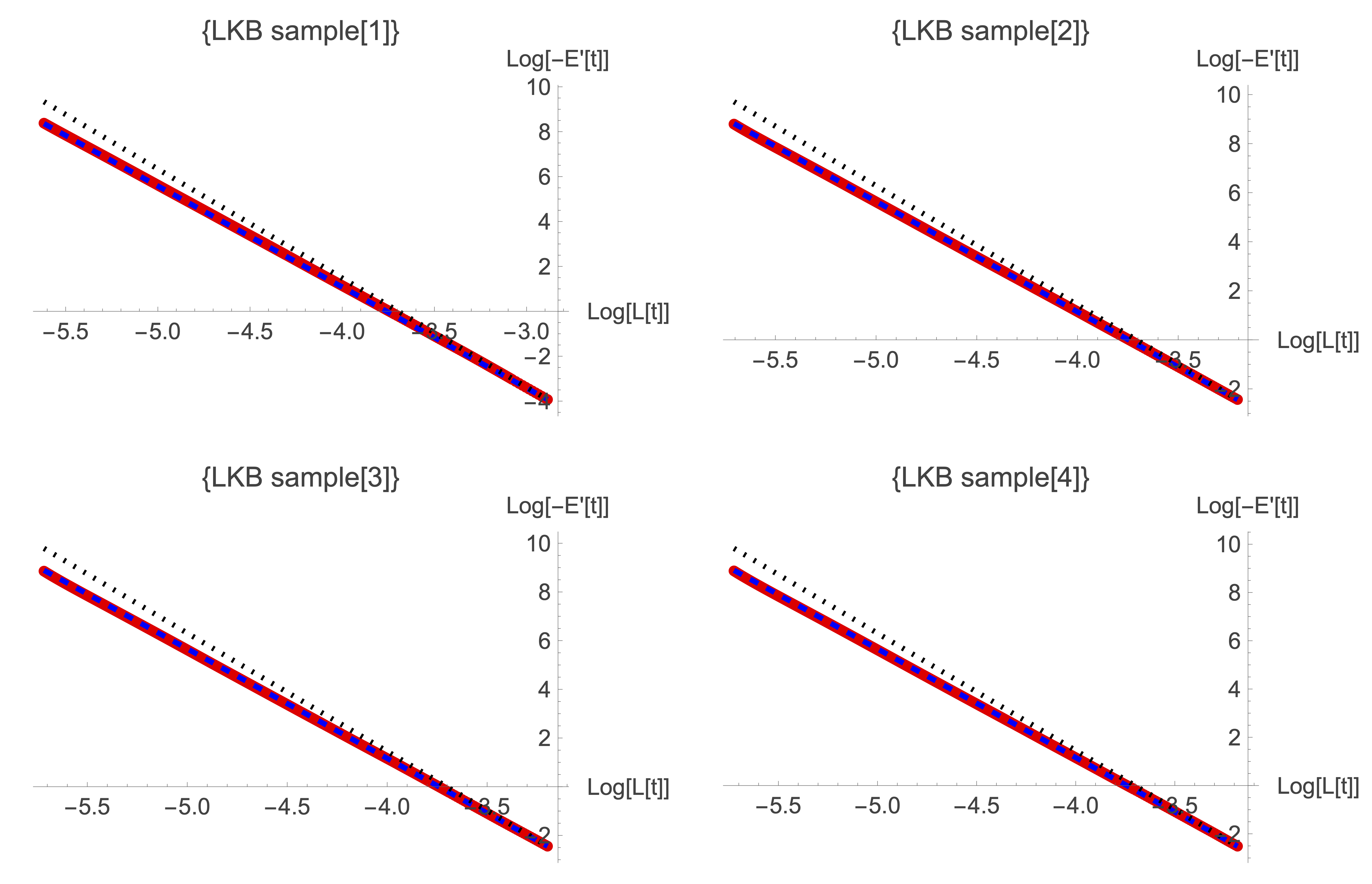}
    \caption{The total enstrophy decay for the LKB regime, when $E(k,0) \propto k^4$. The four plots correspond to four different simulations with randomized initial data.
    The log of enstrophy $\log En(t)$ is plotted against $\log L(t)$ where $L(t)$ is defined as ratio  of first and second moments of $k$ over the whole range of $k$ for a given time in the turbulent region of time moments $ 100 < t < 500$, where $t(L)$ is well approximated by a parabola. The upper limit $500$ was close to the end of available data for the LKB regime. The red dotted line -- the DNS data from  \cite{SreeniAkash2025}, the blue dashed line is our scaling law $En(t) \propto L^{-9/2}$, and the black dotted line corresponds to the K41 law  $En(t) \propto L^{-34/7}$. \textbf{The $9/2$ law fits the data perfectly}, but the $34/7$ is not even close. } 
    \label{fig:LKBEnstrophyDecay}
\end{figure}

We measured the enstrophy decay and compared it with the predicted scaling law.
This comparison is made at Figs.  \ref{fig:LKBEnstrophyDecay}.

This remarkable verification of our scaling law for enstrophy (which is proportional to the time derivative of the energy) suggests that the ostensible violations of universality in the LKB regime for the energy decay were due to contributions from the initial and boundary data.  Once we eliminated these contributions by an extra power of $k^2$ which is present in the enstrophy spectral integral, the universal scaling law of the Euler ensemble was restored.  This resolves the issue of universality violation in the LKB regime as an artifact of the boundary conditions.
\begin{remark}
    Let us clarify the critical issue regarding the universality of the turbulent attractor described by the Euler ensemble and the violation of this universality by initial and boundary data.

    In an infinite system with infinite initial energy, it requires infinite time to dissipate this energy. This scenario corresponds to ideal decaying turbulence, which is described \textbf{exactly} by the Euler ensemble in the limit of zero viscosity at a fixed energy dissipation rate (infinite Reynolds number). This constitutes the mathematical solution to the problem of decaying turbulence.

    However, there is a distinct engineering problem: turbulent flow in a finite system of a particular shape, with finite energy and a finite initial Reynolds number. In such a system, the ideal Euler solution is modified by initial and boundary corrections. The bulk of the decaying energy spectrum in this finite system evolves toward the Euler ensemble solution. This explains why the effective index $\xi_2(r/\sqrt{t})$ in the DNS matches the Euler ensemble prediction perfectly within statistical error. The higher moments of velocity are also calculable in Euler ensemble, though it will take some effort (a good project for an ambitious postdoc, trained in QFT and familiar with mathematical physics and number theory).

    Conversely, the boundary and initial corrections are related to the lower part of the spectrum, which grows into the region of small $k$ within the Euler ensemble solution. In a finite system, this growth is cut off at $k_0 \sim 1/L$, where $L$ is the system size. This infrared part of the spectrum is \emph{not} universal; in Direct Numerical Simulations (DNS), it is strongly influenced by lattice artifacts and finite-size effects. Here, the microscopic theory must step aside, leaving the arena to phenomenological engineering theories such as the K41 scaling laws. These laws reflect dimensional counting and thus apply regardless of the specific dynamics.

    Nevertheless, as the box size increases, these boundary corrections eventually diminish. In the DNS \cite{SreeniAkash2025}, these corrections to the energy decay shift the effective exponents by a few percent. In physical experiments, which operate at much higher Reynolds numbers and in significantly larger domains, the decay index fits our theoretical prediction of $5/4$ perfectly.
\end{remark}

In conclusion, the DNS data \cite{SreeniAkash2025} provide robust validation of our theoretical predictions, agreeing within the statistical and systematic precision (a few percent) of $4096^3$ lattice simulations at Reynolds numbers $\sim 10^2$. This confirmation encompasses the universality of the decay scaling law $E'(t) \propto t^{-9/4}$, the diffusive growth of the effective length $L(t) \propto \sqrt{t}$, and the specific behavior of the effective index (the logarithmic derivative of the second velocity moment as a function of $r/L(t)$). In contrast, alternative phenomenological scaling laws (K41, LKB, BS) fall outside the statistical confidence limits and are thus ruled out. Looking ahead, supercomputing capabilities are poised to surpass $32K^3$ grid resolutions, while wind tunnel experiments are already approaching Reynolds numbers in the tens of thousands. These next-generation studies will be decisive, providing the resolution necessary to disentangle boundary effects from bulk dynamics and to definitively verify the predicted universal attractor.

\subsection{Key Prediction 2: The Geometry of Mixing}
\begin{figure}[h!]
     \centering
     \includegraphics[width=0.8\columnwidth]{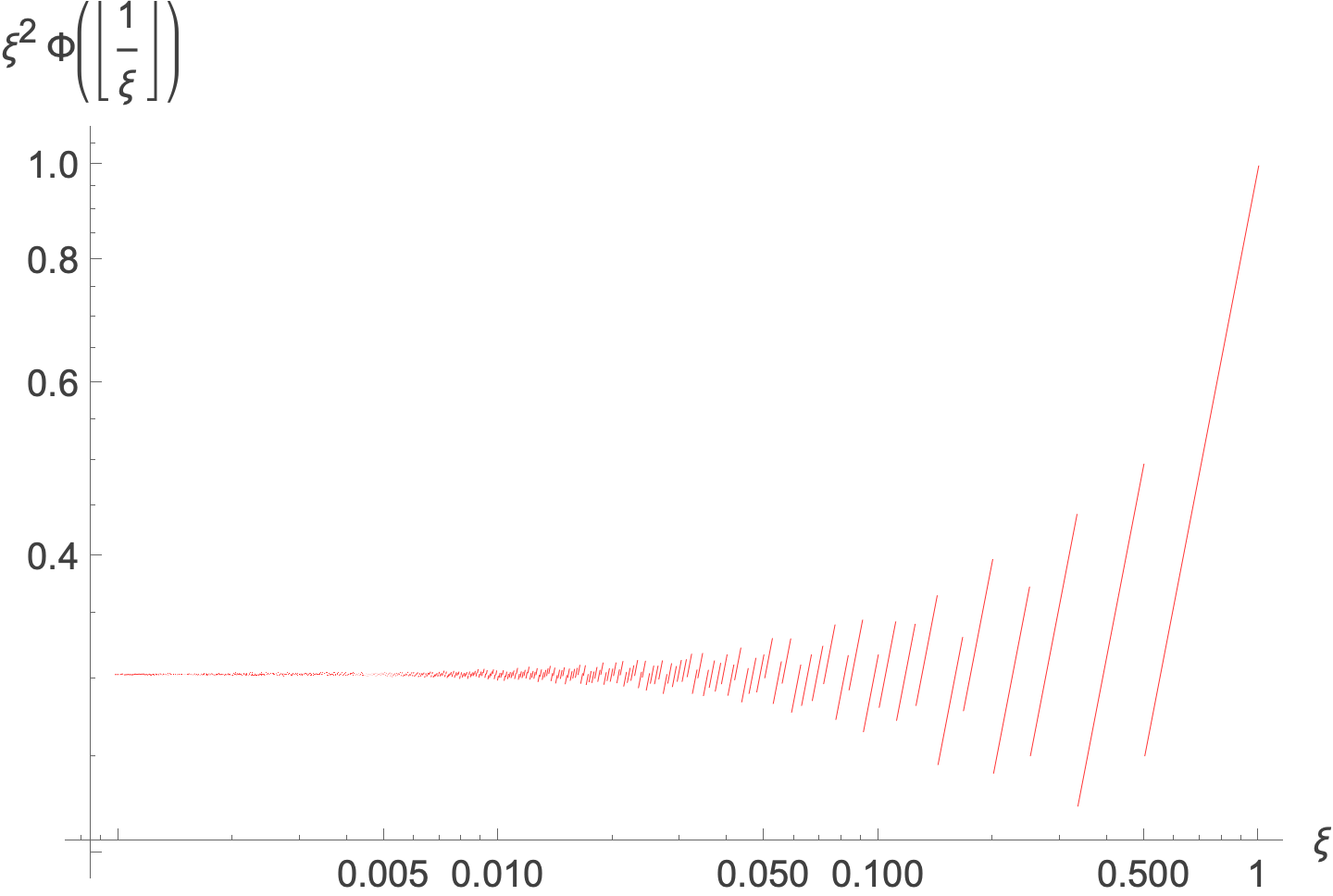}
     \caption{Log--log plot of the universal function $\xi^2\Phi\left(\floor*{\frac{1}{ \xi}}\right)$ where $\xi = \frac{2 \pi r}{\left(\sqrt{2 \tilde \nu (t+ t_0)} -\sqrt{2 \tilde \nu  t_0}\right)}$. Here $\Phi(n) = \sum_{0<p < n} \varphi(p)$ is the Euler totient summatory function.}
     \label{fig::PhiXiPlot}
\end{figure}

The Euler ensemble also dictates the behavior of a passive scalar (e.g., temperature or a dye) advected by the turbulent flow \cite{migdal2025mixing}. For a localized initial condition, the scalar does not spread into a simple Gaussian cloud.
We derived the loop equation for the evolution of the passive scalar $T(r,t)$ from conventional diffusion-advection equations, coupled with the loop equations for the velocity circulation.
The resulting solution for the scalar density in the extreme turbulent limit is quite unexpected.
Instead of a smooth Gaussian cloud, it forms a series of expanding, \textbf{quantized concentric shells} (see Figure \ref{fig::PhiXiPlot}). The radial profile of the scalar concentration is piecewise-parabolic, and the shell structure is organized by the Euler totient function, which counts the number of valid star polygons of a given complexity. This sharp, geometric prediction offers a clear target for future experiments.
\begin{remark}{The velocity singularities vs those of passive scalar}
    Let us clarify the difference between these strong singularities (concentric shells of discontinuities in the passive scalar mean density, condensing to a delta-like singularity in the extreme turbulent limit) and the weaker singularities that the Euler ensemble predicts in velocity correlations. For the velocity field, the two-point correlation function has a first singular term $\delta v^2 \sim r^2 +|r|^{5/2} $, which is weaker  than $|r|^{2/3}$ of K41 model. 
    As for the passive scalar, there are actual infinities and discontinuities generated by the velocity field distributed by the same Euler ensemble.  These singularities may be observed (in some smeared form) as the above ramp-cliffs, but they are clearly impossible to observe in DNS on any finite grid. As we discuss in the research paper submitted to the same issue, the  passive scalar accumulated in a spherical ball around the initial injection point manifests less singularity, but also quite distinctive behavior, which can be observed in DNS. Also, the Fourier transform of the passive scalar density is a smooth universal function of $k\sqrt{t}$, also measurable in DNS.
\end{remark}
\subsection{Key Prediction 3: The Universality of the Energy Spectrum for Arbitrary Space Dimension}

The recent extension of the Euler ensemble theory \cite{migdal2026Zeta} shows that the Euler ensemble \eqref{EulerEnsemble} applies to arbitrary space dimensions $d$, with the last zero component of the vector $\{R \cos \alpha_k,\, R \sin \alpha_k,\, 0\}$ replaced by a zero $(d{-}2)$-dimensional vector and $\hat{\Omega}$ being an $O(d)$ rotation matrix. In particular, \emph{the same Euler ensemble applies to decaying turbulence in two dimensions}, with only two components in the vector $\{R \cos \alpha_k,\, R \sin \alpha_k\}$. A detailed investigation of this universality of the decaying turbulence spectrum and the fine structure of subleading oscillations is presented in the forthcoming paper~\cite{migdal2026Zeta}. The universal energy spectrum in arbitrary dimensions is given by
\begin{align}
   &E(k,t) = \frac{H(k\, L(t))}{L(t)};\quad L(t) = \sqrt{\tilde{\nu}\,(t + t_0)};\label{EspectrumUniv}\\[4pt]
   &H(\kappa) = \int_{-i\infty}^{i\infty} \frac{dp}{2\pi i}\, \kappa^p\, f(p)\,\frac{\zeta\!\left(p + \tfrac{15}{2}\right)\,\Gamma(-p)}{(2p+7)(2p+17)\,\zeta\!\left(p + \tfrac{17}{2}\right)}\,.\label{HkappaUniv}
\end{align}
The universal function $f(p)$, defined and computed in~\cite{migdal2024quantum} using instanton computations for the random walk on regular polygons, is an entire function of the complex variable $p$ independent of the spatial dimension.

The velocity correlation (second moment of velocity difference) depends on dimension~$d$ according to~\eqref{VMellinD}, but the spectrum of intermittency exponents is universal for all space dimensions $d > 1$.

The notorious two-dimensional turbulence has long been viewed as an exception to turbulence theory. In our framework, however, it is not an exception but the simplest manifestation of the universal turbulent attractor based on a quantized momentum loop in the Euler ensemble. The experimental data and DNS for decaying two-dimensional turbulence manifestly contradict the K41 cascade picture but qualitatively agree with our spectrum, including the $k^{-7/2}$ decay law.

We discuss these experiments and the DNS in the forthcoming paper~\cite{migdal2026Zeta}. New DNS based on the particle method (point vortices in 2D) are in progress~\cite{Adolfo2D} and \cite{Sreeni2D}. These simulations are expected to provide the most accurate verification of our theory, given the simplicity of point-vortex dynamics in the planar \NS{} equation.

\section{Application II: Magnetohydrodynamic (MHD) Turbulence}

The loop space solution can be generalized to more complex systems, such as magnetohydrodynamics (MHD), which describes the dynamics of conducting fluids, such as astrophysical plasmas \cite{migdal2025mhd}. The system is now described by two coupled loop functionals, one for the velocity circulation ($\Gamma_v$) and one for the magnetic vector potential circulation ($\Gamma_a$), each governed by an interacting Euler ensemble.

\begin{figure}[h!]
    \centering
    \includegraphics[width=0.8\columnwidth]{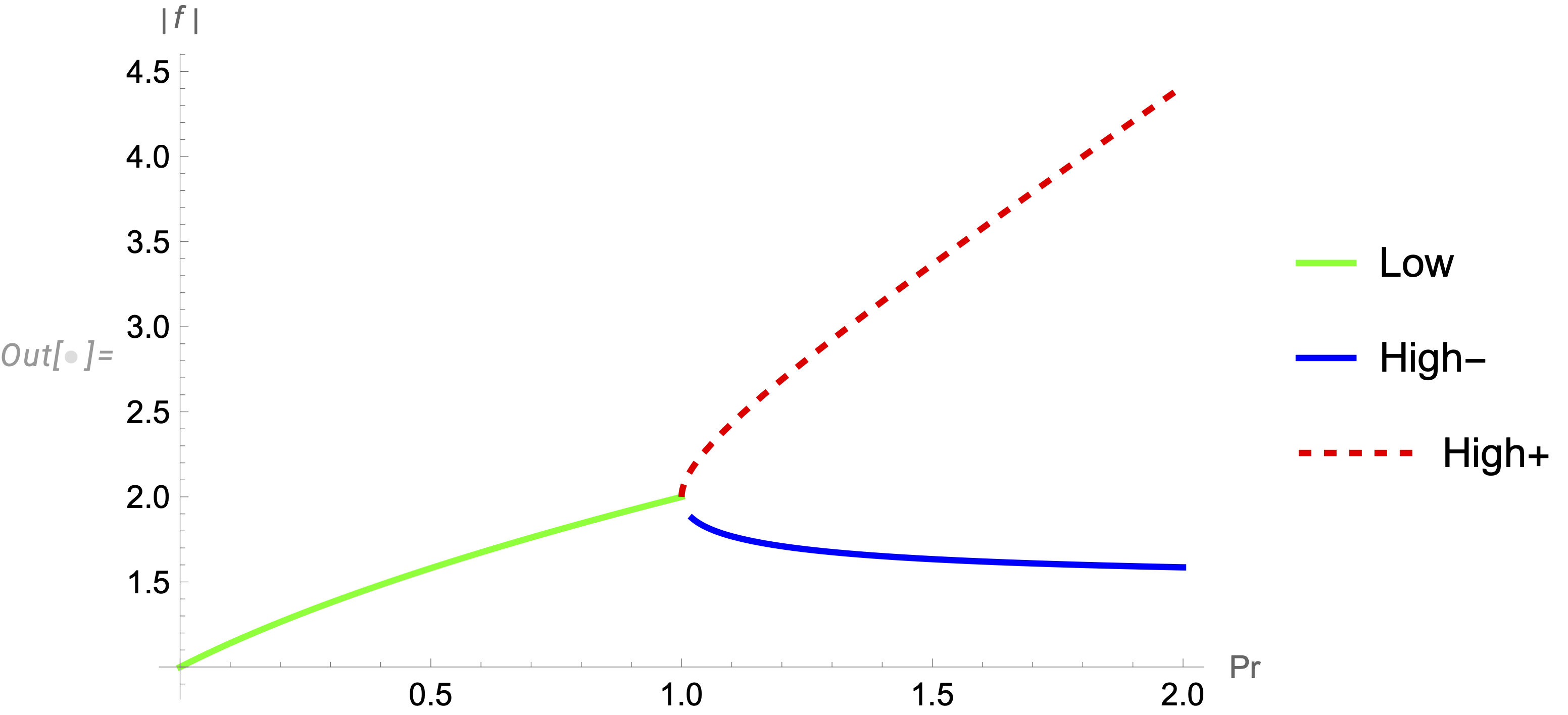}
    \caption{Three phases of the wavevector scale $|f(\Pr)|$ in MHD decaying turbulence. The red-dashed line represents a metastable phase.}
    \label{fig::PrandtlPhases}
\end{figure}

The solution depends critically on the magnetic Prandtl number, $\Pr = \nu/\eta$, the ratio of kinetic viscosity to magnetic resistivity. The theory predicts a \textbf{first-order phase transition at $\Pr = 1$}. (see Figure \ref{fig::PrandtlPhases})
\begin{itemize}
    \item For $\Pr < 1$ (e.g., liquid metals), the turbulence is dominated by hydrodynamic fluctuations.
    \item For $\Pr > 1$ (e.g., hot plasmas), the system exhibits two branches: a stable regime where magnetic and kinetic energies are balanced, and a metastable regime characterized by growing magnetic fluctuations.
\end{itemize}
This sharp, falsifiable prediction provides new insight into the behavior of astrophysical and laboratory plasmas. The scale $|f(\Pr)|$ of the wave vector in the spectrum is the following function of the Prandtl number
\begin{align}
    & |f(\Pr)|  = 
    \begin{cases}
        \sqrt{1 + 3 \Pr} & \text{ if } \Pr < 1\\
        1 + \Pr -\sqrt{\Pr^2 -\Pr }& \text{ if } \Pr > 1
    \end{cases}
\end{align}
\section{The Yang-Mills Connection and Universal Duality}

The final step reveals the framework's profound universality. The Yang-Mills gradient flow, a central tool for understanding the vacuum structure and confinement problem in QCD, is governed by a non-Abelian version of the same loop dynamics \cite{migdal2025YangMills}. Applying our framework, we find exact solutions for the evolution of the Wilson loop.

In the Yang--Mills gradient flow, the loop-space calculus leads not only to the
diffusion equation for the Wilson loop but also to its stationary fixed point,
which defines the confining string.  The analytic solution is provided by the
\emph{Hodge-dual matrix surface}, obtained by the harmonic-map method.  This
surface satisfies the self-duality relation
\begin{align}
\Sigma_{\mu\nu}=\sfrac{1}{2}\epsilon_{\mu\nu\alpha\beta}\Sigma_{\alpha\beta},
\end{align}
and minimizes the dual area functional under this constraint.
The surface area is defined as 
\begin{align}
\label{areaDef}
    &S_\chi[C] = \min_{X}\int_{\Sigma} d^2 \xi\sqrt{\Sigma_{\mu\nu}^2/2};
 \end{align} 
 The area element is defined as
\begin{align}
\label{sigmaDef}
    \Sigma_{\mu\nu} = \epsilon_{a b}\partial_a X^i_\mu\partial_b X^i_\nu;
 \end{align} 
 The Hodge chirality $\chi = \pm 1$ enters through the boundary conditions
    \begin{align}
    \label{BCX}
        &X^i_\mu(\partial \Sigma) = \eta^{\chi,i}_{\mu\nu} C_\nu ;
    \end{align}
Here $\eta^{\chi,i}_{\mu\nu}$ are 't Hooft's matrices corresponding to Hodge duality $\chi = \pm 1$
    \begin{align}
      \eta^{\chi,i}_{\mu\nu} =  \delta_{i\mu} \delta_{\nu 4} - \delta_{i\nu}\delta_{\mu 4}  + \chi e_{i\mu\nu 4} ;
    \end{align}
Its area  $S[C]$ is an exact zero mode of the loop diffusion operator in four dimesions
\begin{align}
    \mathcal L S[C] =0;
\end{align}
And as a consequence,there is a symmetry in the non-perturbative fixed-point Yang--Mills
loop equation (so called Makeenko-Migdal loop equation \cite{MM1981NPB, Mig83}),
\begin{align}
&\mathcal L(W[C]) = \lambda \int_C d x_\mu \int_C d y_\mu\delta^4(x-y) W[C_{xy}]W[C_{yx}],
\end{align}
ensuring that $\exp{-\kappa S[C]}$ is annihilated by the loop operator.  This
construction yields a parity-even, factorized form of the Wilson functional,
\begin{align}
\label{confiningFactor}
W[C]=W_0[C]\exp{-\kappa(S_{+}[C]+S_{-}[C])},
\end{align}
where $S_{\pm}[C]$ are the areas of the analytic self-dual and anti-self-dual
surfaces bounded by the same loop, and $W_0[C]$ is the perturbative solution corresponding to asymptotic freedom.
The minimality conditions with the extra self-duality constraint allow an exact harmonic solution inside the unit disk ($|z| <1$)
\begin{align}
    &X^a_\mu(z,\bar z) = \eta^{\chi,i}_{\mu\nu} (f_\nu (z) + \bar f_\nu(\bar z));\\
    & C_\mu(\theta) = 2 \Re f_\nu \left(e^{\I \theta}\right)
\end{align}
For non-intersecting loops, the dual area is $S_\pm[C]=2\sqrt2|A[C]|$, where $A[C]$ is the minimal Euclidean area bounded by $C$. This reproduces the confining area law, while for self-intersecting loops, our solution is additive across the closed parts, unlike the Euclidean minimal surface, which switches between topologies to achieve the global minimum of the area.  Our additive solution, however, represents the local minimum of the area, which is sufficient for the confining factor to satisfy the MM loop equations. 

This analytic result provides the
first explicit geometric realization of the Yang--Mills confining string and
establishes a bridge between gauge theory, dual string geometry, and the
loop-space formulation of turbulence.

The fact that the very same quantized number-theoretic structure governs the universal attractor in both classical fluid turbulence and fundamental gauge theory is the strongest evidence of a deep unifying principle. This duality—between a strongly coupled, chaotic field theory and a weakly coupled, solvable string theory on a discrete target space—is a new manifestation of a principle familiar from the AdS/CFT correspondence.
\section{Suggestions for Future Experiments and Tests of the Theory}

The loop space calculus framework provides a rich set of new, quantitative, and falsifiable predictions that can be addressed by the next generation of experiments and high-resolution direct numerical simulations (DNS). We outline two primary directions for future work that could provide definitive tests of the theory.

\subsection{Visualizing Quantized Scalar Shells}
\begin{figure}[htbp]
  \centering
  \includegraphics[width=0.4\textwidth]{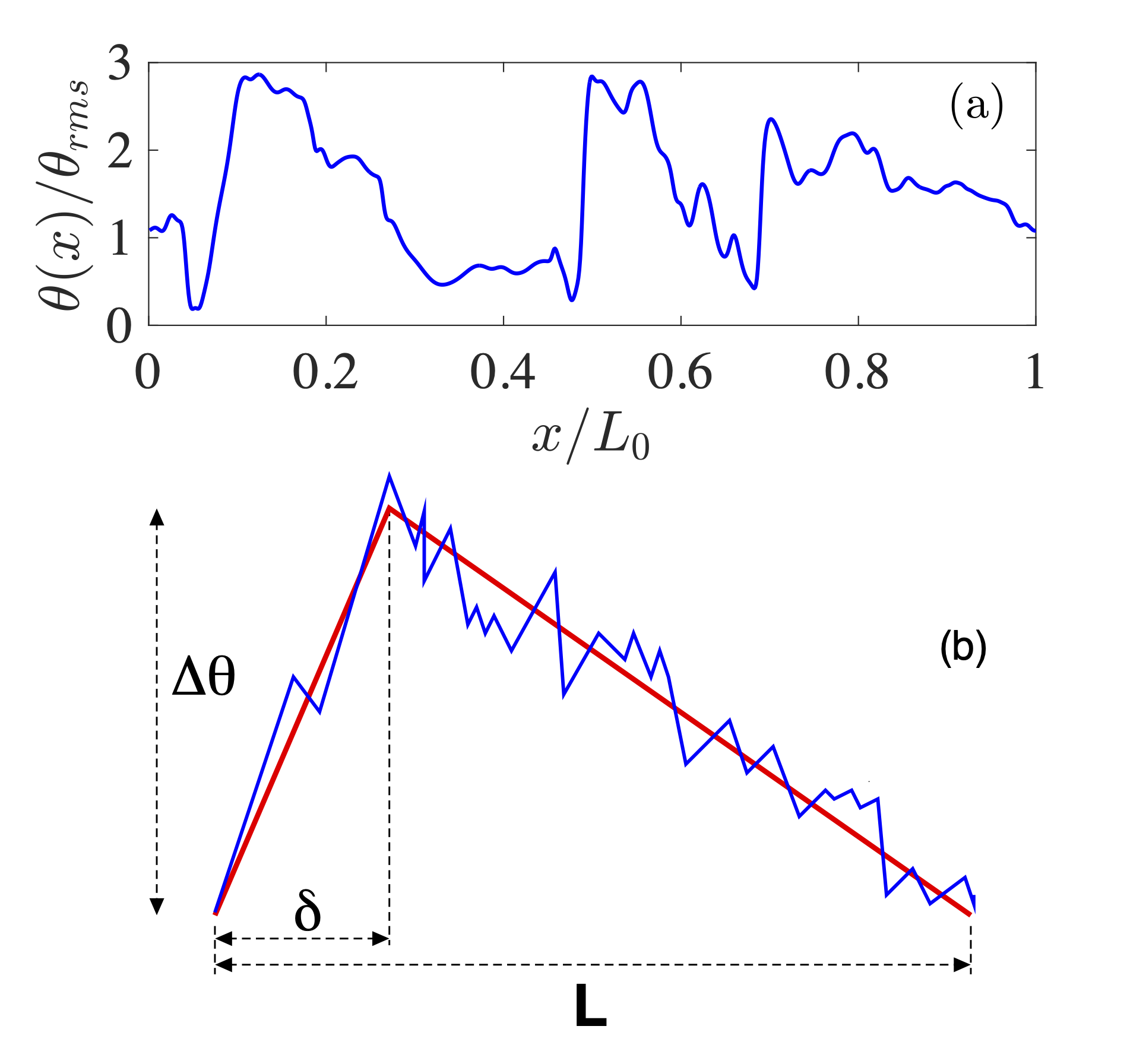}
  \caption{The 'ramp-cliff' structure in the time trace of temperature fluctuations ($\Delta\theta$) in a heated turbulent jet. These asymmetric patterns, characterized by a gradual rise (the 'ramp') followed by a sharp drop (the 'cliff'), are a key signature of large-scale coherent structures imprinting on small-scale scalar fields. This empirical observation is analogous to the sawtooth profile predicted for the quantized scalar shells in the Euler ensemble (Figure \ref{fig::PhiXiPlot}) if you reflect this picture across the right-hand side.( Figure adapted from \cite{sreenivasan2018turbulent}}
  \label{fig:ramp-cliffs}
\end{figure}
As detailed in Section 4(b), our theory predicts that a passive scalar released from a localized source in decaying isotropic turbulence will not simply diffuse into a Gaussian cloud. Instead, it should form a series of discrete, expanding concentric shells, whose spacing and density profile are determined by the Euler totient summatory function. The time-domain signature of these shells at a fixed sensor would be a characteristic "sawtooth" decay (see Figure \ref{fig::PhiXiPlot}).

While existing DNS data has been invaluable for testing spectral properties, most studies have focused on fluctuation statistics rather than the coherent evolution of the mean field. A definitive test of this unique spatial prediction requires a dedicated numerical experiment. To this end, a collaboration has been initiated to solve the passive-scalar advection-diffusion equation concurrently with an existing high-resolution DNS of decaying isotropic turbulence. This project will provide the first direct test for the existence of these quantized shell structures. The singular space density would require very large computer resources, currently unachievable, but the less singular integrated density inside  a sphere of given radius, or, equivalentlty, the amplitude of monochromatic wave with a given wavevector, are also measurable and these are smooth functions, detectable on a large grid DNS. 

Presumably, these shell structures manifest as the "ramp-cliff" structures with similar sawtooth profiles, observed long ago but never completely understood from the microscopic theory ( see \cite{sreenivasan2018turbulent}, (see Figure \ref{fig:ramp-cliffs})

\subsection{Direct measurement of the Mellin transform in two and three dimensions}

The newly discovered universality of decaying turbulence across spatial dimensions offers a unique opportunity to verify the Euler ensemble solution using particle methods in two-dimensional vortex dynamics. The most direct test would be the universal scaling law $\sqrt{\tilde{\nu} t} E(k,t) = H(k \sqrt{\tilde{\nu} t})$ within the turbulent range of wavevectors ($k$ in the bulk, bounded by the inverse system size from below and the lattice upper cutoff from above), evaluated at times large enough to erase the memory of initial data, yet sufficiently early to maintain a large, time-decaying Reynolds number.

\begin{figure}[htbp]
    \centering
    \includegraphics[width=0.9\textwidth]{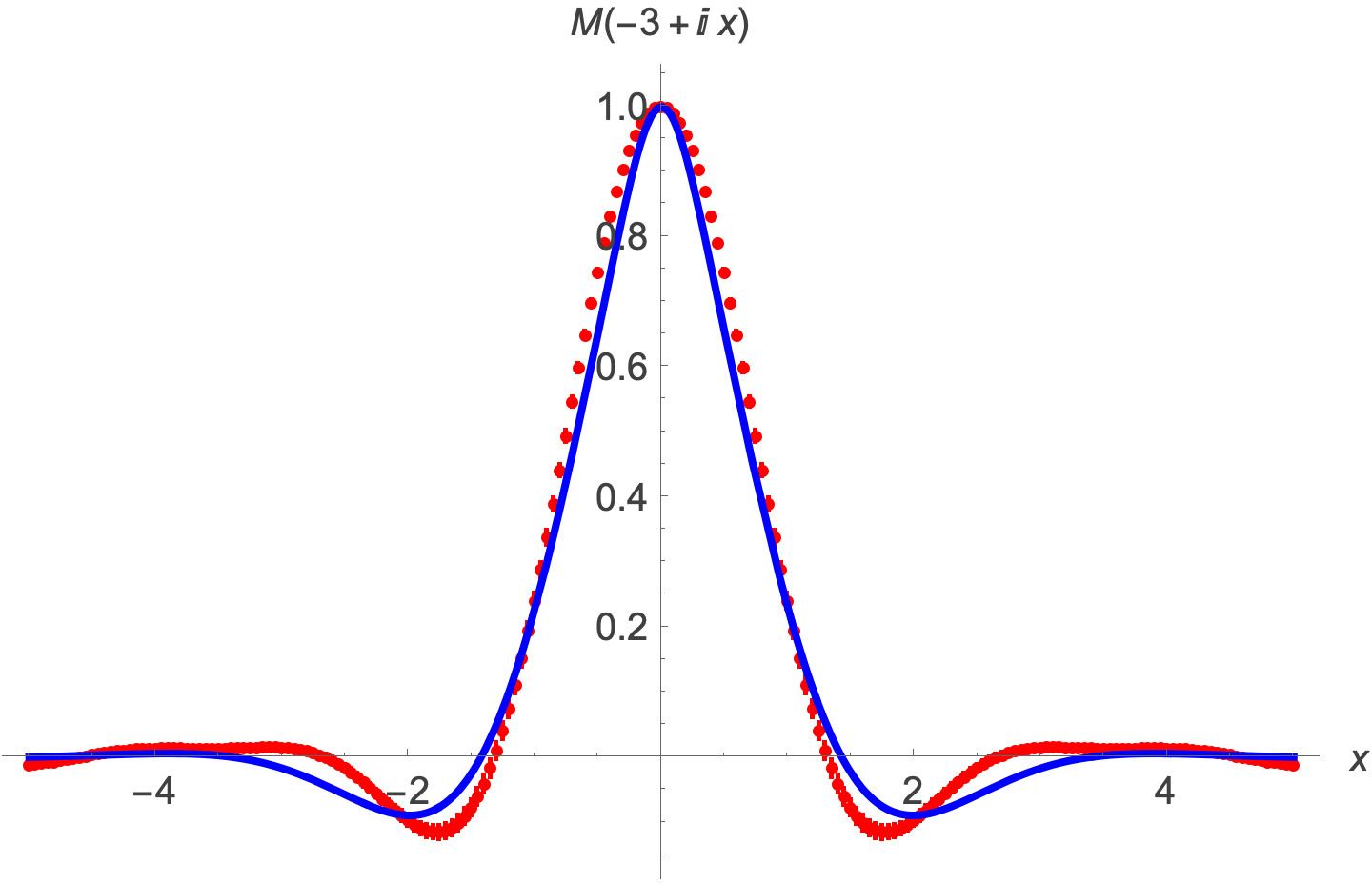}
    \caption{The real part of the complex Mellin transform evaluated along the contour $p=-3+\I x$, normalized such that $\Re[M(-3)]=1$ and $\Re[M^{\prime}(-3)]=0$. The pure, parameter-free Euler loop theory is shown as the solid blue curve. The Mellin values, extracted from time averages of spectral data from 3D DNS \cite{SreeniAkash2025} at a fixed scaling variable $k L(t)$, are shown as red markers. The vertical error bars represent the standard error ($\pm 1\sigma/\sqrt{N}$) of the temporal fluctuations, assuming uncorrelated deviations from the Euler ensemble. In reality, due to intermittency, correlations exist that effectively increase these error estimates.}
    \label{fig:RealMellin}
\end{figure}

The $k^{-7/2}$ decay law provides the next critical target, but the entire shape of the universal scaling function $H(\kappa)$ predicted by our theory for intermediate $\kappa$ is also directly measurable. This is analogous to measuring the effective index of the velocity moment, as was done above in 3D. A closely related bulk observable is the Mellin transform $\int_0^\infty d\kappa \, H(\kappa)/\kappa^{p+1}$ evaluated at complex arguments $p = -3 + \I x$ (for $|x| < 5$), where the spectral integral safely converges. This Mellin transform is an explicit meromorphic function of the complex variable $p$, and it closely matches the 3D DNS results \cite{SreeniAkash2025}. We present this comparison in Figure \ref{fig:RealMellin}, previewing results from a forthcoming paper \cite{migdal2026Zeta}.

\section{Beyond Kolmogorov: The Limits of Classical Cascade Theories}\label{discussion}

\subsection*{Grid Turbulence and DNS}
\begin{figure}[htbp]
    \centering
    \includegraphics[width=0.48\columnwidth]{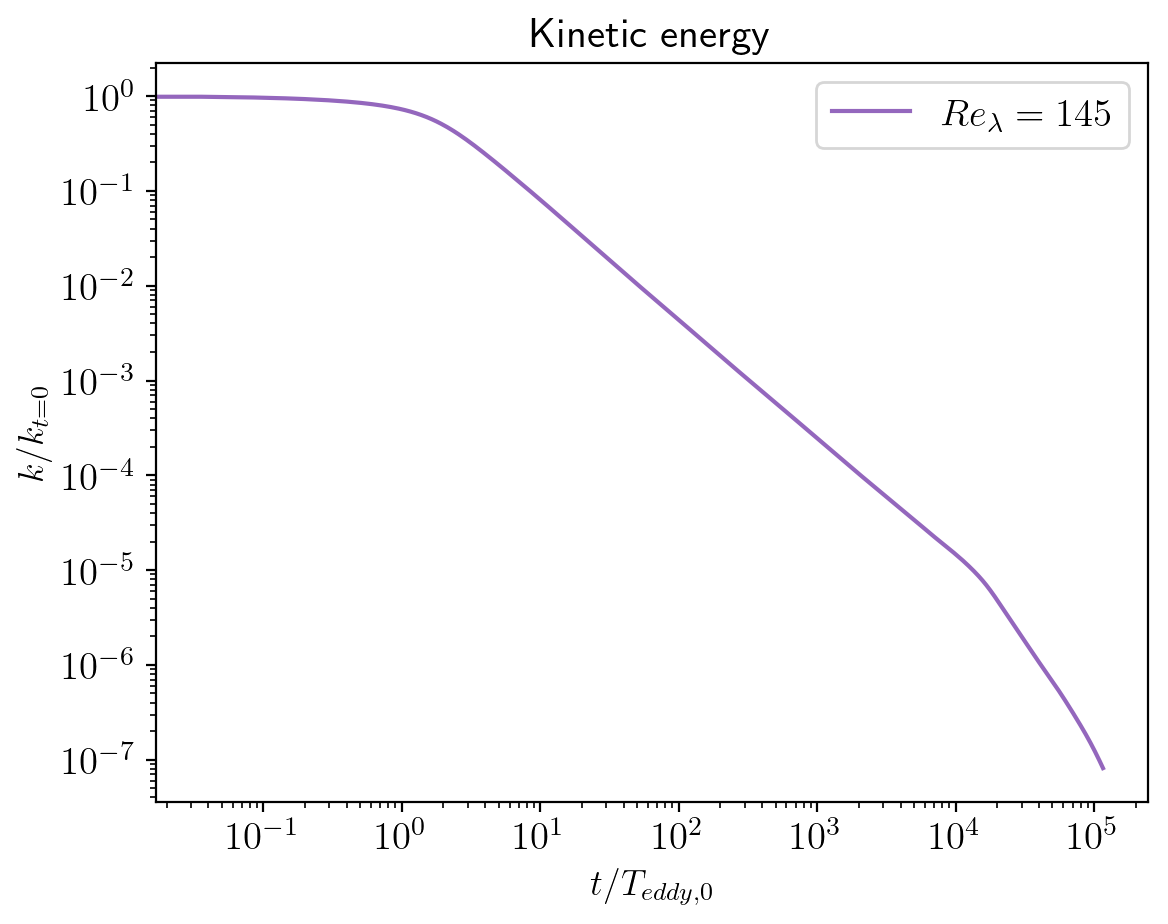}
    \includegraphics[width=0.48\columnwidth]{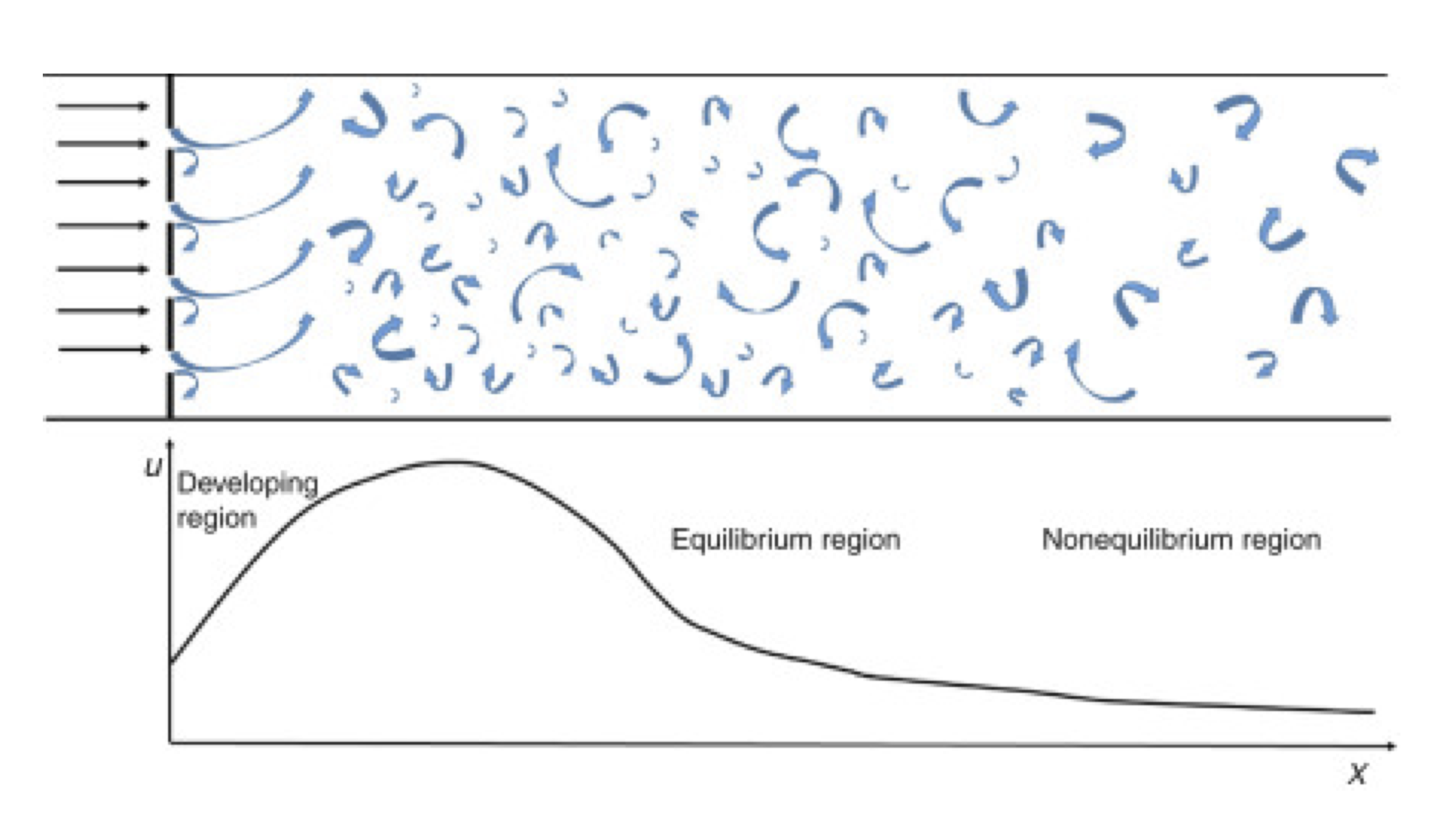}
    \caption{Left: The turbulent kinetic energy decaying with time in the $4096^3$ grid simulation with initial Gaussian velocity field distributed according to the K41 spectrum $k^{-5/3}$. After a few hundred time steps, the statistical equilibrium was reached, and energy started decaying in agreement with our $t^{-5/4}$ law. There are other parameters of that stage of decay, with distribution matching our theory, such as the length scale $L(t) \sim \sqrt{t}$ and effective index of the second moment of velocity field which is a nontrivial function of scaling variable $\log (r/\sqrt{t})$ (see Fig.~\ref{fig:DNSFit}). Right: Typical setup of the grid turbulence experiment. The flow enters from the left, passes through the oscillating grid, and creates vorticity. These vortexes interact, exchange energy, and eventually reach statistical equilibrium, with energy constantly dissipated in the bulk while being resupplied by the flow through the grid. In the still Galilean frame, this is a steady process, with energy coming from the left boundary and dissipated in the bulk. The local kinetic energy decays with the distance $z$ from the grid as $z^{-1.25}$, but stays constant with time. In the frame moving with the mean flow velocity, this is decaying turbulence, with energy decaying with time as $(v t)^{-1.25}$.}
    \label{fig:GridTurbulenceExperiment}
\end{figure}

A classic setting is grid turbulence (Fig.~\ref{fig:GridTurbulenceExperiment}).
The landmark experiments of \cite{GridTurbulence_1966} measured a decay exponent
$1.25\pm 0.01$, close to $5/4$ and inconsistent with the Kolmogorov–Saffman value
$6/5$. More recently, DNS on $4\mathrm K$ grids with periodic boundaries and K41-type initial spectra (no forcing) were performed by Sreenivasan and collaborators. The
simulations relaxed to a decaying state in which the energy decays as $t^{-5/4}$, the
integral scale obeys $L(t)\sim t^{1/2}$, and the effective index of the second
moment follows the predicted nonlinear curve (Fig.~\ref{fig:GridTurbulenceExperiment},Left),
\ref{fig:DNSFit}).
\subsection*{Relation to Classical Turbulence Paradigms}
The loop equations as a route to decaying turbulence represent a departure from
customary practice, and they naturally meet resistance from the community. The
objection is often framed as follows: rather than resolving the myriad vortical
structures and their interactions in physical space, the loop approach reduces the problem to the solution of a stochastic process on regular star polygons
(Fig.~\ref{fig:RegularPolygons}). Can such a one-dimensional description be truly
equivalent to three–dimensional vortex dynamics?

A related concern is that any putative solution must address the familiar themes of
the classical picture: energy cascade, vortical hierarchies, multifractal scaling,
and intermittency. Where do these ideas sit in the loop description?
Our position is simple. Some widely used assumptions and scaling templates do not
apply to the \emph{decaying} turbulence regime; others are modified or generalized in the loop
picture. Below, we explain how the main questions are answered within the theory.

\subsection*{The Physical Interpretation of Loop Space Geometry}
The polygons in Figure  \ref{fig:RegularPolygons} are not meant to be seen in a flow visualization; they are a convenient
basis in \emph{loop space}. Their role is analogous to an unobservable but predictive
degrees of freedom in other areas of physics (such as quark confinement): they lead to quantitative, testable
predictions, such as the decay of turbulent kinetic energy as $t^{-\frac{5}{4}}$.

\subsection*{Duality and the Recasting of Complexity}

The key is \emph{duality}: the same physics can admit complementary mathematical descriptions. In one representation, the dynamics look intricate (waves and eddies); in another, they are encoded by a simpler stochastic object. Just as strong coupling in one theory can map to weak coupling in a dual theory, the turbulent limit of the velocity dynamics maps to a WKB-like limit of the Euler ensemble, which is exactly solvable at infinite Reynolds number. In this sense, the “complexity” is not lost; it is recast. 

The complexity is thus not diminished, but rather transformed into a different mathematical language. It shifts from the statistical phenomenology of multifractal exponents and moment-closure hierarchies to the deep arithmetic structures of number theory, revealing a connection between turbulence and concepts such as Euler totients and the Riemann zeta function.

\subsection*{The Nature of the String Theory Duality}

A natural question arises from the perspective of string theory: In what sense is the Euler ensemble a "string theory" if the loop $C$ is not the boundary of a worldsheet?

The connection is one of duality, analogous to the relationship between position and momentum space in quantum field theory. The Euler ensemble solution for the loop functional, Eq.~\eqref{Psisol}, is precisely the generating functional of a 1D quantum field theory—a string theory—living on the loop itself.

Let us break down the analogy:
\begin{itemize}
    \item \textbf{The String Field:} The momentum loop, $f_\mu(\theta)$, which executes a random walk on a discrete target space (the vertices of star polygons), plays the role of the string's position field.
    
    \item \textbf{The Source/Momentum:} The physical loop, $\dot{C}_\mu(\theta)$, does not represent the string itself. Instead, it acts as a \textit{source} or an external momentum field that couples to the string's position, $f_\mu(\theta)$.
    
    \item \textbf{The Amplitude:} The loop functional $\Psi[C]$ is, therefore, the string's vacuum-to-vacuum amplitude in the presence of this external momentum source. Averaging over the Euler ensemble is the path integral over all allowed string configurations.
\end{itemize}

This correspondence manifests a \emph{momentum-space duality}. The loop $C$ lives in our physical space, but it acts as a momentum-space probe for a string residing in a separate, dual space.

This interpretation elegantly resolves the apparent paradox of a discrete spectrum. The quantization and discreteness (the star polygons, the number-theoretic structure) exist in the dual momentum space of the string. The physical loops $C$ in our world remain continuous and smooth. The theory does not assume nor predict discrete loops in physical space, but rather a discrete spectrum of statistical correlations that are probed by continuous loops.

Thus, the Euler ensemble is not a conventional string theory where the loop is a boundary. It is a dual string theory with a discrete target space, where the physical loop $C$ functions as the momentum variable in its generating functional.
\subsection*{The Hodge--Dual Surface as a Theoretical Area Law in QCD}

Why introduce a new area law when lattice QCD already confirms the conventional one?
The answer lies in distinguishing a numerical observation from an analytic solution.
Two arguments are central.

\begin{itemize}
\item \textbf{Analytic foundation.}  Numerical simulations can test or falsify models but cannot
derive the law itself from first principles.  The analytic Hodge--dual matrix surface,
constructed by the harmonic--map method, exactly solves the fixed--point Yang--Mills loop
equation \cite{migdal2025geometric}.  It provides the theoretical explanation for the confining area law rather than an
empirical fit, offering a genuine analytic counterpart to the numerical area law of lattice QCD.

\item \textbf{Beyond current lattice reach.}  Present lattice resolutions are insufficient to test the
distinctive, shape--dependent predictions of the Hodge--dual surface.  Existing calculations
mostly involve planar Wilson loops, for which the dual area is $2\sqrt2$ times the Euclidean
area---a factor that can be absorbed into the string tension.  The crucial test would
involve nonplanar, ``twisted'' contours where the dual and Euclidean minimal surfaces yield
different areas.  Such computations remain demanding but would provide a definitive check
of the analytic theory once feasible.
\end{itemize}

In this sense, the Hodge--dual matrix surface does not replace the lattice area law but explains
its origin and predicts measurable deviations for nonplanar geometries, establishing an analytic
framework for the QCD confining string. Being an exact analytic solution, it also provides a
natural starting point for quantizing the string, beginning from this holomorphic minimal-surface
solution. The resulting quantized theory offers a systematic path to computing the hadron
spectrum from first principles, extending far beyond the phenomenology of the linearly rising
potential.

\subsection*{Kolmogorov Scaling Violation in the Decaying Regime}
With respect to the decaying regime, the classical K41 template is not supported by
data at the level often assumed. Kolmogorov’s original constant–dissipation ansatz
was a model assumption, and even he (with Obukhov) introduced fluctuations
(log–normal) soon after. Modern experiments and DNS
\cite{YZ93,YS04,SY21} show significant, systematic deviations from K41 (and from its
log–normal variant). In decaying turbulence, experiments \cite{SreeniDecaying} deviate
even more strongly, despite attempts to fit or trim the data; for example, the energy
spectrum departs from $k^{-\frac{5}{3}}$ over many decades, and recent measurements
\cite{GregXi2} report clear departures of the "effective index" of $\langle\Delta v^2\rangle(r)$
from the K41 value of $2/3$.

\subsection*{Multifractals and Conformal Symmetry}
Motivated by analogies with critical phenomena, multifractal models posit power laws
with anomalous exponents in an inertial range and can organize deviations from K41 in
forced turbulence \cite{FP85,YS04,SY21}. In the incompressible turbulence, however, there is no
theoretical basis for conformal symmetry: the conservation constraints of a CFT would
assign \emph{equal} dimensions $d-1=2$  to divergence-less fields $\vect v , \vect \omega$. 
This relation is incompatible with $\boldsymbol\omega=\nabla\times\mathbf v$, which would lead to \emph{dimensions differing by $1$} rather than being equal. 
More importantly, the decaying spectra are \emph{not} straight lines on log–log plots; they
bend universally (cf. Fig.~2 (top) of \cite{GregXi2}), indicating that single
power laws—whether K41 or multifractal—are insufficient. The loop theory predicts
nonlinear, universal functions of $\log k\sqrt{t}$ across several decades, consistent with
the observed curvature.

\subsection*{A Generalization of Multifractal Laws}
While not conformally invariant, loop space theory provides a microscopic basis for and a generalization of the multifractal paradigm: the multifractal models introduce a continuous spectrum of anomalous exponents phenomenologically, but the loop theory derives, from first principles, an infinite, discrete, and arithmetic spectrum of intermittency exponents.
$$ 
\text{CFT: }\vec v(0) \cdot \vec v(r) \sim \sum_n O_n |r|^{p_n};\; p_n \stackrel{?}{=} 2 \Delta_v - \Delta_n;
$$
The Mellin transform of the velocity correlation is found to be meromorphic, yielding the set of scaling exponents shown below:
\begin{table}[h]
\centering
\renewcommand{\arraystretch}{1.3}
\begin{tabular}{l c}
\hline\hline
\textbf{Pole position} & \textbf{Multiplicity} \\
\hline
$2n \quad (n \in \mathbb{Z}, n\ge0)$ & 1 \\[2pt]
$\sfrac{5}{2}$ & 1 \\[2pt]
$\sfrac{11}{2}$ & 1 \\[2pt]
$\sfrac{15}{2} + 2n \quad (n \in \mathbb{Z},\; n \geq 0)$ & 1 \\[2pt]
$7 \pm \I\,\rho_n \quad (n \in \mathbb{Z}_{\geq 1})$ & 1\textsuperscript{*} \\[2pt]
\hline\hline
\end{tabular}
\caption{Poles of the Mellin kernel $V(p)$ with positive real part. Here $\rho_n$ denotes the imaginary part of the $n$-th nontrivial zero of the Riemann zeta function, $\zeta(\tfrac{1}{2} + i\rho_n) = 0$. The asterisk indicates that the multiplicity is~1 assuming the Riemann hypothesis (i.e., that all nontrivial zeros are simple).}
\label{tab:poles}
\end{table}
where \( \frac{1}{2} \pm \imath \rho_n \) are the nontrivial zeros of the Riemann zeta function.

This result fundamentally refines the multifractal idea. The leading exponents in this spectrum, consistent with numerical data, can be viewed as the "ground state" of the scaling law, describing the dominant behavior. The infinite series of subsequent exponents then dictates a hierarchy of calculable, quantum-like corrections. Recent DNS~\cite{DecayingTalk} report slopes consistent with these predictions, in contrast to the classical scaling templates of Kolmogorov or Saffman.
\subsection*{The 4/5 Law, Spectral Flux, and Vorticity}

A central pillar of the classical cascade narrative is the Kolmogorov 4/5 law. In its familiar scalar form, the law relates the third-order moment of the longitudinal velocity increment to the separation distance $r$, i.e., $\langle (\delta v_L)^3 \rangle = -\frac{4}{5}\epsilon r$. This linear dependence on $r$ is widely interpreted as direct evidence for a constant, scale-independent energy flux through the inertial range.

A more complete analysis, however, requires examining the law's full tensorial structure, $S_3^{\alpha\beta\gamma}(\mathbf{r}) = \langle v_\alpha(\mathbf{0}) v_\beta(\mathbf{0}) (v_\gamma(\mathbf{r}) - v_\gamma(\mathbf{0})) \rangle$. The exact solution to the Karman-Howarth equation shows that this tensor is \textbf{strictly linear} in the separation \textbf{vector} $\mathbf{r}$ rather than its length $|\mathbf{r}|$.

This linearity has a direct and crucial consequence in Fourier space. Its Fourier transform is proportional to the \textbf{gradient of a delta function}, $\nabla\delta(\mathbf{k})$, meaning its entire physical content is localized to the largest scale of the system ($\mathbf{k}=0$). The law, therefore, provides a rigorous constraint on large-scale dynamics but does not describe a process of energy transfer \textit{between} scales.

Furthermore, this tensor is fundamentally \textbf{irrotational}. The law is therefore silent on the statistics of \textbf{vorticity}, which is the essential ingredient of the turbulent cascade. Indeed, the most direct third-order mixed correlation involving vorticity, $\langle v_\alpha(\mathbf{0}) v_\beta(\mathbf{0})  \omega_\gamma(\mathbf{r}) \rangle$, \textbf{vanishes identically} in isotropic turbulence for purely kinematical reasons of parity invariance. The 4/5 law, thus, offers no constraint on the non-trivial vorticity correlations that would be required to describe a scale-to-scale energy transfer, nor does it limit the scaling properties of vorticity correlations.

\subsection*{The Vanishing Flux and the Dissipative Anomaly}

The present framework calls into question not only the K41 scaling laws but also the conventional concept of a "Kolmogorov energy flux." The rapid decay of the energy spectrum predicted by the theory, in accordance with bounds established by Sulem and Frisch~\cite{Sulem_Frisch_1975}, corresponds to a vanishing Kolmogorov flux in the turbulent limit. The apparent paradox of finite dissipation in the absence of a spectral flux is resolved by the dissipative anomaly, a mechanism not accounted for in that earlier work.

This mechanism suggests an alternative physical picture to the Richardson cascade. Rather than a scale-by-scale transfer, energy injected at the largest scales (e.g., from boundaries) is dissipated directly on fine-scale, viscous micro-structures. The exact expressions for dissipation and the energy spectrum derived from the loop space approach~\cite{migdal2024quantum} are consistent with the dissipative anomaly and this picture of direct dissipation, but not with a constant-flux energy cascade.

\subsection*{Heisenberg's Dissipative Range Model}
Heisenberg’s $k^{-7}$ proposal for a dissipative subrange, taken up by Chandrasekhar,
was a model of its time \cite{Heisenberg1948PRSA,Chandrasekhar1949PRSA}.
As Sreenivasan recounts \cite{SreeniChandra}, von Neumann already noted in 1949 the
lack of experimental support. In our Mellin spectrum, there is no pole between
$-13/2$ and $-8\pm\mathrm i t_n$, and the "effective index" approaches $-7/2$ rather
than $-7$. In other words, the nontrivial dynamics persist across what used to be
separated as “inertial” and “dissipative” ranges. The resulting universal decaying
spectrum spans several decades, and the "effective index" of $\langle\Delta v^2\rangle(r)$
varies smoothly from $2$ down to $0$, with no plateau at $2/3$. These predictions
match DNS and experiments across the entire turbulent range without adjustable
dimensionless parameters.

\subsection*{The Question of Universality and Relevance to Steady Flows}
The applicability of results from forced simulations to universal turbulence characteristics requires careful consideration. With finite forcing kept active, the resulting statistical state can depend significantly on the specifics of the forcing scheme (e.g., its spatial structure, correlation time, and the mechanism of energy injection), potentially violating universality (see, e.g., the discussion in \cite{Ishihara2009}). In particular, large-wavelength forcing, especially if targeting the potential (irrotational) velocity component, can potentially seed non-universal large–scale structures influenced by long–range correlations $\langle v v \rangle(r) \sim r^{\alpha} \to \infty$. In such cases, the simulated flow may reflect properties of the driving mechanism more than the intrinsic fluid dynamics, akin to altering the taste of soup with the greasy spoon used to stir it.

By contrast, the loop equation isolates the vorticity sector, whose correlations inherently decay with distance and are expected to be less sensitive to boundary conditions and to details of distant forcing. This separation helps identify a universal decaying–turbulence attractor. Crucially, this attractor is directly relevant even to ostensibly \emph{steady} turbulent flows, such as Feynman's pipe flow, experimental grid turbulence, and jets. These flows typically exhibit spatial decay downstream from the generation source. By Galilean invariance, in a frame moving with the mean flow velocity $v$, this spatial decay is equivalent to the temporal decay problem addressed by our theory, with downstream distance $z$ corresponding to time $t = z/v$ and the boundary condition at the energy source, such as the inlet of the tube, becoming an initial condition at $t=0$ when a given element of water passes the energy source. The observed spatial energy decay in grid turbulence, $E(z) \sim z^{-5/4}$ \cite{GridTurbulence_1966}, provides a compelling example, matching the temporal $E(t) \sim t^{-5/4}$ decay predicted by the Euler ensemble. Thus, the study of temporally decaying turbulence offers a first-principles approach to understanding the universal aspects of these physically ubiquitous, spatially developing flows, addressing the long-standing challenges highlighted by Feynman.

\subsection*{Ergodicity on the Attractor}

In Newtonian mechanics, the evolution of the probability distribution in phase space (the Liouville equation $ \partial_t \rho = \{H, \rho\} $) is solved by any function of the Hamiltonian $H$. Physical arguments select the Gibbs distribution, $\rho \propto \exp{-\beta H}$. This argument relies on the properties of the stationary solution itself, not directly on the ergodic hypothesis—the (still unproven) conjecture that a trajectory of a conservative Hamiltonian system eventually covers its energy surface uniformly. Despite the lack of a general proof for ergodicity, the Gibbs distribution has been successfully applied for over a century.

In our theory, the loop diffusion equation replaces the Liouville equation, and the fixed-point (decaying) solution that replaces the Gibbs distribution is the Euler ensemble. The analog of the ergodic hypothesis is the conjecture that every distinct state of the Euler ensemble contributes equally to long-time averages, implying that each state is visited with equal frequency over time. We have not proven this conjecture; the closest approach involved obtaining a transcendental equation for the spectrum of the time decay of deviations from the Euler ensemble \cite{migdal2023exact}. A proof of ergodicity (or convergence to the attractor) would need to demonstrate the stability of the Euler ensemble, i.e., the positivity of the real parts of all the decay indices in its spectrum.

Lacking such a proof, we turn to physical and numerical experiments. These consistently show convergence from initial states with large vorticity towards a solution whose statistical properties closely match the predictions derived from the Euler ensemble, including decay indices and spectra, as discussed in previous sections \cite{GridTurbulence_1966, TurbulentDecaySreeni22, SreeniAkash2025}. While the stability and ergodic properties of the Euler ensemble remain a mathematical challenge, this empirical agreement provides strong support for its role as the relevant invariant measure for decaying turbulence.
\begin{remark}
    One of the thoughtful readers of this manuscript asked the following question:
    \begin{quote}
        Finally, considering the quantum and linear nature of the diffusion-loop equation, does that author have any comments on the possibility of utilizing quantum computers to study these systems? ...
    \end{quote}
    This question resonates deeply with our own experience. During the long struggle with the Navier-Stokes loop equation—before the analytic simplicity and beauty of the Euler ensemble emerged—we explored various numerical avenues. Even now, after two years of successfully investigating the properties of the Euler ensemble in infinite space, the necessity for a powerful numerical method remains acute for \emph{finite} geometries.
    
    Quantum computing indeed offers a unique opportunity here. The chance lies not in verifying the Euler ensemble (which is exact for the infinite case), but in solving the loop diffusion (Schrödinger) equation in a \emph{finite geometry} by mapping it onto a quantum mechanical system of qubits. This approach appears far more efficient than the existing proposals for mapping the nonlinear Navier-Stokes equation onto a grid of qubits. In the case of the loop equation, we possess a linear Hamiltonian structure (albeit non-Hermitian) from the outset. The qubits would live on a loop, moving in a physical space, which looks like a huge reduction of dimensionality.
    
    We have made several passes towards experts in quantum simulations, only to find that, for the moment, such simulations remain at the level of science fiction. However, a breakthrough in quantum hardware could occur at any time. Therefore, we believe it is vital to keep this unique path open: solving for large-scale turbulent flows around physical bodies by simulating the loop equation on a future quantum computer.
\end{remark}

\subsection*{Summary.}
In the decaying turbulence, single–power scaling laws and cascade arguments do not
describe the observed behavior. The loop equation replaces these templates with a
microscopic, parameter–free description in arbitrary space dimension $d >1$: universal nonlinear spectral shapes,
an arithmetic spectrum of indices, and a diffusion in loop space that organizes the
statistics. The evidence from classic experiments and modern DNS supports this
picture across the turbulent range.
\section{Conclusion and Outlook}

The loop space calculus offers a new perspective on a class of strongly nonlinear problems in theoretical physics. The approach is based on reformulating the dynamics from local fields to non-local loop observables, recasting the governing nonlinear equations into a universal linear diffusion equation in the space of loops. This transformation enables direct analytical treatment by applying the functional Fourier transform, leading to a solvable algebraic equation.

For decaying turbulence in arbitrary space dimension, this has led to a parameter-free solution---the Euler ensemble---which is shown to be dual to a solvable string theory. A central result is the discovery of a deep connection between the flow's spatial statistics and its temporal evolution. Where phenomenological models treat them separately, this framework derives both the spectrum of spatial \textbf{intermittency exponents} and the spectrum of temporal \textbf{decay exponents} from a single, underlying number-theoretic structure related to the zeros of the Riemann zeta function. The leading exponents in these spectra correspond to observed scaling and decay laws, while the infinite series of subsequent exponents provides a hierarchy of calculable, quantum-like corrections. The theory's predictions are consistent with available numerical data and offer clear targets for future experimental tests.

The appearance of the same mathematical solutions in the context of the Yang-Mills gradient flow suggests a broad applicability of these methods. These applicability points towards a common structure underlying the statistical behavior of these disparate physical systems. Promising directions for future research include applying this approach to problems in compressible turbulence, cosmology, and the loop dynamics of Einstein gravity.

Among these avenues, the theory's formulation of the confining string is of particular interest for quantum chromodynamics. The framework yields an exact solution to the QCD loop equations in the form of a \textbf{Hodge-dual minimal surface}~\cite{migdal2025YangMills}. This geometric object, by its very construction, satisfies the loop equations in four-dimensional space, unlike the conventional Euclidean minimal surface, which violates the full set of \YM{} loop equations. 

The analytic Hodge--dual matrix surface derived by the harmonic--map
method~\cite{migdal2025geometric} provides the geometric core of
the Yang--Mills fixed--point solution, completing the dual correspondence between turbulence and confinement.

This explicit formula for the shape dependence of the quark string evolution opens the way to computing the rising Regge trajectories in QCD
from first principles \cite{Migdal2026GeometricQCDII, migdal2026GeometricQCDIII}.

While the same mathematical formalism connects turbulence, gauge theory, and string duality, these correspondences should not be interpreted as a unification of the underlying physical systems  but rather as a shared geometric language linking universal features of their non-linear dynamics.

\medskip 

\medskip

Sometimes, understanding familiar physical phenomena requires learning new mathematical methods, as happened historically with planetary motion and chemical reactions. In my pursuit of a solution to turbulence, I found it necessary, over decades of exploration, to learn relevant aspects of geometry and number theory. This process eventually led to a geometric understanding of loop-space diffusion and the parallel transport operator: the loop-space calculus. This new perspective then naturally revealed the turbulent attractor rooted in number theory—the Euler ensemble. Assuming that this solution withstands further scrutiny and validation, it invites a shared effort to explore this rich mathematical landscape, echoing the spirit of Euler, who brought together distinct fields of mathematics to describe natural laws.

\section*{Acknowledgements}
I am grateful for insightful discussions regarding this theory with participants of seminars and conferences at the International Centre for Theoretical Sciences (ICTS) in Bangalore (India), where this theory was first reported, the Kavli Institute for Theoretical Physics (KITP), the Perimeter Institute for Theoretical Physics, Rutgers University, the Institute for Advanced Study (IAS), the University of Oxford, the Isaac Newton Institute for Mathematical Sciences in Cambridge, the London Institute for Mathematical Sciences, Imperial College London, the CERN Theoretical Physics Department, the "Turbulence on the Banks of the Arno" workshop in Pisa (2025), the 2nd European Fluid Dynamics Conference (EFDC2) in Dublin (2025), and numerous Zoom seminars in the US and Europe. Conversations with Katepalli Sreenivasan concerning the physics of turbulence, as observed in direct numerical simulations and laboratory experiments, were invaluable in clarifying the role and significance of loop space diffusion.

Supported by the Simons Foundation (2019–2024).

\bibliographystyle{plainnat}
\bibliography{bibliography} 

\appendix

\section{Essentials of the loop space calculus}
\label{sec:appendix_calculus}
\addcontentsline{toc}{section}{Essentials of the loop space calculus}

The \loopcalculus{} is a variational framework that operates entirely within the manifold of smooth loops. It replaces singular operations (e.g., adding infinitesimal loops) with coincident limits of smooth "dot derivatives" $\ff{\dot C(\theta)}$. This appendix summarizes the key definitions, identities, and results of the calculus.

The full derivations and technical details can be found in our recent paper \cite{migdal2025SQYMflow}.

\subsection*{The Wilson loop and non-abelian Stokes theorem}
The area derivative is defined by the non-abelian Stokes theorem.
For a detailed derivation, see \cite{migdal2025SQYMflow}, Sec. 3.3.
\begin{align} \label{eq:A_stokes}
    \delta W[C] = \oint d \theta  \dot C_\mu(\theta) \delta C_{\nu}(\theta) \fbyf{W[C]}{\sigma_{\mu\nu}(\theta)}
\end{align}

\subsection{Operator representation of the Wilson loop}
The loop functional can be represented as a trace of a path-ordered exponential of covariant derivative operator. This subsection proves the fundamental operator identity that enables this.
The full derivation, originally presented in \cite{migdal2025SQYMflow} (Sec. 3.4), is reproduced here with all necessary mathematical details. We also fix some typos in that paper.

\begin{theorem}[Operator-Holonomy Identity]
Let $D_\mu(x) = \mathbb I_G \partial_\mu + A_\mu(x)$ be the covariant derivative at an arbitrary base point $x \in \mathbb{R}^d$, where $\partial_\mu$ is the partial derivative in Hilbert space and $A_\mu(x)$ is a matrix-valued operator in the group space $G$, and $\mathbb I_G$ is a unity matrix in that group space. Let $C$ be a smooth, closed loop parameterized by $\theta \in [0, 2\pi]$ with $\dot{C}_\mu(\theta)$ as its velocity, such that $C(\theta) = x + \int_0^\theta d\theta' \dot{C}_\mu(\theta')$, and $\int_0^{2\pi} d\theta' \dot{C}_\mu(\theta') = 0$.

The path-ordered exponential of the covariant derivative operator $D_\mu(x)$ along this loop is equal to the Wilson loop (holonomy) along the path $C(\theta)$ multiplied by the identity operator $\mathbb{I}$ in the Hilbert space.
\begin{align}
    \mathbb{P} \exp{ \int_0^{2\pi} d\theta \dot{C}_\mu(\theta) D_\mu(x) } = 
    \mathbb{P} \exp{ \int_0^{2\pi} d\theta \dot{C}_\mu(\theta) A_\mu(C(\theta)) } \otimes \mathbb{I}
\end{align}
\end{theorem}

\begin{proof}
The proof relies on discretizing the path and using the "disentangling" identity, which is a consequence of Feynman's operator calculus \cite{Feynman1951}.

\paragraph{1. Discretization}
The path-ordered exponential on the L.H.S. is formally defined as the limit of a product integral:
\begin{align}
    \mathbb{P} \exp{ \int_0^{2\pi} d\theta \dot{C}_\mu(\theta) D_\mu(x) } = 
    \lim_{N\to\infty} \prod_{k=N \to 1} \exp{ \Delta\theta_k \dot{C}_\mu(\theta_k) D_\mu(x) }
\end{align}
where $\Delta\theta_k \to 0$ and the product is ordered from right to left (i.e., $k=1$ is the rightmost operator). Let $dC_\mu(\theta) = d\theta \dot{C}_\mu(\theta)$.

\paragraph{2. The Infinitesimal Disentangling}
We analyze a single infinitesimal factor $\exp{dC_\mu D_\mu(x)}$. Using the Lie-Trotter product formula, $e^{A+B} = e^A e^B + O([A,B])$, we can split the operator:
\begin{align}
    \exp{ dC_\mu (\partial_\mu + A_\mu(x)) } = 
    \exp{ dC_\mu \partial_\mu } \exp{ dC_\mu A_\mu(x) } + O(d\theta^2)
\end{align}
The $O(d\theta^2)$ terms vanish in the $N\to\infty$ limit. The $\exp{dC_\mu \partial_\mu}$ operator is an infinitesimal translation operator, $T_{dC}$. We now use the fundamental operator identity that defines a translation:
\begin{align}
    T_{dC} \, f(x) = f(x+dC) \, T_{dC}
\end{align}
Applying this to our $A_\mu(x)$ operator, we find the "hopping" identity:
\begin{align}
    \exp{ dC_\mu \partial_\mu } \exp{ dC_\mu A_\mu(x) } = 
    \exp{ dC_\mu A_\mu(x+dC) } \exp{ dC_\mu \partial_\mu }
\end{align}
This is the "disentangling" from (A7) in the text. It allows us to move the (Abelian) derivative operator to the right, at the cost of shifting the argument of the (non-Abelian) $A_\mu$ operator.

\paragraph{3. Iteration and Re-ordering}
We now apply this identity iteratively to the full product. Let $C_k = C(\theta_k) = x + \int_0^{\theta_k} d\theta' \dot{C}(\theta')$.
By periodicity, $ C_N = x$.
\begin{align}
    &\prod_{k=N \to 1} \exp{ dC_k \cdot D(x) } \approx \prod_{k=N \to 1} \left( \exp{ dC_k \cdot A(C_k) } \exp{ dC_k \cdot \partial_x } \right) \\
    &= \left( e^{dC_N \cdot A(C_N)} e^{dC_N \cdot \partial_x} \right) \left( e^{dC_{N-1} \cdot A(C_{N})} e^{dC_{N-1} \cdot \partial_x} \right) \dots \left( e^{dC_1 \cdot A(C_N)} e^{dC_1 \cdot \partial_x} \right)
\end{align}
We repeatedly "hop" all the $\exp{dC_k \cdot \partial_x}$ terms to the far right. Each time a $\partial_x$ operator passes an $A(C_N)$ operator, it shifts its argument. The crucial point is that all $\partial_x$ operators are Abelian and commute with each other, while all $A(C_k)$ operators are non-Abelian and remain path-ordered.
Furthermore, the resulting argument $x_k$ of each $A$ in the product at $k-th$ place will be
\begin{align}
    x_k = C_N + \sum_{n=N \to k} dC_{n} = C_k
\end{align}
This process separates the product into two distinct, ordered parts:
\begin{align}
    = \left( \prod_{k=N \to 1} e^{dC_k \cdot A(C_k)} \right) \left( \prod_{k=N \to 1} e^{dC_k \cdot \partial_x} \right)
\end{align}
Taking the $N\to\infty$ limit, this becomes:
\begin{align}
    = \left( \mathbb{P} \exp{\int_0^{2\pi} d\theta \dot{C}_\mu(\theta) A_\mu(C(\theta))} \right) \left( \exp{\int_0^{2\pi} dC_\mu \partial_\mu} \right)
\end{align}

\paragraph{4. The Closed Loop Condition}
The first term is, by definition, the Wilson loop $W[C]$ (the holonomy), which is a c-number matrix. The second term is the identity operator in Hilbert space, because the loop $C$ is closed:
\begin{align}
    \int_0^{2\pi} dC_\mu = \int_0^{2\pi} d\theta \dot{C}_\mu(\theta) = C_\mu(2\pi) - C_\mu(0) = 0
\end{align}
Therefore, the operator becomes:
\begin{align}
    \exp{ \left(\int_0^{ 2 \pi}d C_\mu(\theta)\right)\pd{x_\mu} } = \exp{ 0 \cdot \pd{x_\mu} } = \mathbb{I}
\end{align}
This completes the proof. We find that the Hilbert space operator on the L.H.S. is equal to the group space matrix (the Wilson loop) multiplied by the identity operator $\mathbb{I}$.
\end{proof}

\subsection*{The area derivative}
The field strength $F_{\mu\nu}$ is generated by the antisymmetric part of the second dot derivative.
For details, see \cite{migdal2025SQYMflow}, Sec. 3.6, Eq. (25).
\begin{align} \label{eq:A_areader}
    &\fbyf{W(C(.),\tau)}{\sigma_{\mu\nu}(t)} = \ff{\dot C_{\lb{\mu}}(t-)}\ff{\dot C_{\rb{\nu}}(t+)}W(C(.),\tau)  = \nonumber\\
    &\VEV{\frac{1}{N}\tr F_{\mu\nu}(C(t)) P \exp{\I \int_t^{t+2\pi} d s \dot C_\mu(\theta) D_\mu(C(t))}}
\end{align}

\subsection*{The covariant derivative}
The covariant derivative of the field strength is generated by the discontinuity of the dot derivative, $\partial_\mu(t) \equiv \ff{\dot C_\mu(t-)}-\ff{\dot C_\mu(t+)}$.
For details, see \cite{migdal2025SQYMflow}, Sec. 3.7, Eq. (31).
\begin{align} \label{eq:A_covDer}
    &\partial_\mu(t) \fbyf{W[C]}{\sigma_{\mu\nu}(t)} = \VEV{\tr \left[D_\mu, F_{\mu\nu}(C(t))\right]\right.\nonumber\\
    &\left.P \exp{\I \int_t^{t+2\pi} d \theta \dot C_\mu(\theta) D_\mu(C(t))}}
\end{align}

\subsection*{Annihilation property}
The dot derivative discontinuity $\partial_\mu(t)$ annihilates the Wilson loop.
For details, see \cite{migdal2025SQYMflow}, Sec. 3.9, Eq. (33).
\begin{align} \label{eq:A_annihil}
    \partial_\mu(t) W(C(.),\tau) =0;
\end{align}

\subsection*{The Leibniz rules}
The area derivative acts as a first-order differential operator, satisfying the standard Leibniz rules.
For details, see \cite{migdal2025SQYMflow}, Sec. 3.10, Eqs. (39-41).
\begin{align}
    \ff{\sigma_{\mu\nu}(\theta)} (A[C] B[C]) &= \fbyf{A[C]}{\sigma_{\mu\nu} (\theta)} B[C] + A[C] \fbyf{B[C]}{\sigma_{\mu\nu} (\theta)}; \label{eq:A_Leibniz1} \\
    \ff{\sigma_{\mu\nu}(\theta)} F(A[C]) &= F'(A[C]) \fbyf{A[C]}{\sigma_{\mu\nu} (\theta)}; \label{eq:A_Leibniz2} \\
    \pd{\mu} \ff{\sigma_{\mu\nu}(\theta)} (A[C] B[C]) &= 0, \quad \text{if A, B solve the fixed-point eq.} \label{eq:A_Leibniz3}
\end{align}

\subsection*{The Bianchi identity}
The calculus contains a kinematical Bianchi identity.
For details, see \cite{migdal2025SQYMflow}, Sec. 3.11, Eq. (43).
\begin{align} \label{eq:A_ident}
    e_{\alpha\mu\nu\lambda} \left(\ff{\dot C_\alpha(\theta+)} - \ff{\dot C_\alpha(\theta-)}\right) \ff{\dot C_{\lb{\nu}}(\theta-)} \ff{\dot C_{\rb{\mu}}(\theta+)} =0;
\end{align}

\subsection*{The nonsingular loop equation}
The dot derivatives combine to form the loop operator $\mathcal{L}_C$, yielding a closed, linear diffusion equation for the Wilson loop functional.
For details, see \cite{migdal2025SQYMflow}, Sec. 3.12, Eqs. (45-48).
\begin{align}
    \partial_{\tau}W[C,\tau] &= \mathcal{L}_{C}W[C,\tau] \label{eq:A_loop_eq} \\
    \mathcal{L}_{C} &= \oint d\theta \dot C_{\nu}(\theta)\hat{L}_{\nu}(\theta); \label{eq:A_loop_op} \\
    \hat{L}_{\nu}(\theta) &= T_{\nu}^{\alpha\beta\gamma}\frac{\delta^{3}}{\delta\dot{C}_{\alpha}(\theta-0)\delta\dot{C}_{\beta}(\theta)\delta\dot{C}_{\gamma}(\theta+0)}; \label{eq:A_L_hat} \\
    T_{\nu}^{\alpha\beta\gamma} &= \delta_{\alpha\beta}\delta_{\gamma\nu}+\delta_{\gamma\beta}\delta_{\alpha\nu}-2\delta_{\alpha\gamma}\delta_{\beta\nu}; \label{eq:A_TensorT}
\end{align}
\hypertarget{sub-the-nonsingular-loop-equation}{}
\section{Back to the static loop equation}
\setcounter{equation}{0}
\renewcommand{\theequation}{B\,\arabic{equation}}

We now map the Euler-ensemble solution of the liquid loop equation back to the static
loop equation by prescribing a loop motion in momentum space that reproduces the
missing $\,\mathbf v\times\boldsymbol\omega\,$ term. Consider
\begin{align}
     &\Psi[C,t] = \Big\langle \exp{\I\oint d\theta\ \mathbf C'(\theta)\cdot \mathbf P(\theta)}\Big\rangle_{\mathcal E},\\
     & \partial_t \mathbf C(\theta) = -\,\mathbf v_P(\theta),\\
     & \frac{D}{Dt}\Psi[C,t] = \partial_t\Psi[C,t]\nonumber\\
     &     - \Big\langle \I\oint d\theta\ \partial_\theta \mathbf v_P(\theta)\cdot \mathbf P(\theta)\,
           \exp{\I\oint \mathbf C'\cdot \mathbf P}\Big\rangle_{\mathcal E}.
\end{align}
Integrating by parts and using $\I\,\partial_\theta\mathbf P\Rightarrow-\,\delta/\delta\mathbf C$
on the exponential yields the desired static advection operator. 
The velocity is expressed in terms of vorticity by inverting the curl, given zero divergence.
\begin{align}
     \vec v = \frac{-1}{\vec \nabla^2} \vec \nabla \times \vec \omega
\end{align}
With the dictionary (valid inside the momentum loop Anzatz)
\begin{align}\label{OpRel}
     & \nabla \;\Rightarrow\; \frac{\partial}{\partial \mathbf C(\theta)}
           = \frac{\delta}{\delta\mathbf C'(\theta-0)} - \frac{\delta}{\delta\mathbf C'(\theta+0)},\\
      & \hat\omega(\theta) \;\Rightarrow\; -\,\I\,\nu\,\frac{\delta}{\delta\boldsymbol\sigma(\theta)}\nonumber\\
       &
           = -\,\I\,\nu\,\frac{\delta}{\delta\mathbf C'(\theta-0)}
              \times \frac{\delta}{\delta\mathbf C'(\theta+0)}.
\end{align}
we find
\begin{align}
     & \nabla \Rightarrow \frac{\delta}{\delta\mathbf C'(\theta-0)}-\frac{\delta}{\delta\mathbf C'(\theta+0)} \Rightarrow -\,\I\,\Delta\mathbf P,\\
     & \boldsymbol\omega \Rightarrow
       \I \nu \frac{\delta}{\delta\mathbf C'(\theta-0)}\times\frac{\delta}{\delta\mathbf C'(\theta+0)}\nonumber\\
       &
          \Rightarrow \I \nu \mathbf P_{\mathrm{mid}}\times \Delta \mathbf P,
\end{align}
so that
\begin{align}
     \mathbf v_P = \nu\frac{\Delta \mathbf P \times (\mathbf P_{\mathrm{mid}}\times \Delta \mathbf P)}{(\Delta \mathbf P)^2}
     \quad\stackrel{\mathcal E}{\Longrightarrow}\quad
     \mathbf v_P = \nu \mathbf P_{\mathrm{mid}}.
\end{align}
The extra term in momentum space is then
\begin{align}
   & \I\oint d\theta\ \partial_\theta \mathbf v_P\cdot \mathbf P
        = -\,\I\oint d\theta\ \mathbf v_P\cdot \partial_\theta \mathbf P\nonumber\\
       &
        = \I\nu\oint d\theta\ \mathbf P_{\mathrm{mid}}\cdot \partial_\theta \mathbf P,
\end{align}
which vanishes: the smooth part $\mathbf P_{\mathrm{mid}}$ of $\mathbf P$ gives an integral of total derivative $\frac{1}{2} \partial_\theta \mathbf P_{\mathrm{mid}}^2$ over the closed loop, and the
jump  contributions $\mathbf P_{\mathrm{mid}}\cdot \Delta\mathbf P$ vanish locally by the Euler-ensemble constraints. Equivalently, solving
for the moving momentum loop
\begin{align}
     \mathbf C(\theta,t) = \mathbf C_0(\theta)
           + \big(\sqrt{2\nu(t+t_0)}-\sqrt{2\nu t_0}\big)\,\mathbf F_{\mathrm{mid}}(\theta)
\end{align}
one finds that the time-dependent term cancels as in the liquid equation. Thus, the Euler ensemble solves both the liquid and static loop equations.
\section{Velocity Correlations and the Spectrum of Exponents}
\setcounter{equation}{0}
\renewcommand{\theequation}{C\,\arabic{equation}}

The Euler ensemble solution yields an explicit formula for the second moment of the velocity difference as a Mellin transform:
\begin{equation}
    \langle (\Delta \vect{v})^2 \rangle(r,t) = \frac{\tilde{\nu}^2}{\nu t} \int_{\epsilon - i\infty}^{\epsilon + i\infty} \frac{dp}{2\pi i}\, V(p) \left(\frac{r}{\sqrt{\tilde\nu\, t}}\right)^p
\end{equation}
where $\tilde{\nu}$ is a turbulent viscosity, and the kernel $V(p)$ is a meromorphic function given by:
\begin{equation}
   V(p) =\frac{\Gamma \left(\frac{d-1}{2}\right) f(-p-1)\, \zeta \left(\frac{13}{2}-p\right) \sec \left(\frac{\pi(p+1)}{2}\right) \Gamma \left(\frac{p+1}{2}\right)}{32 \pi ^2 (2 p-15)(2 p-5)\, \zeta \left(\frac{15}{2}-p\right) \Gamma \left(\frac{d+p}{2}\right)}
   \label{VMellinD}
\end{equation}
Here, $d = 2, 3, \ldots$ is the dimension of space, $\zeta(s)$ is the Riemann zeta function, and $f(z)$ is a calculable entire function defined in Appendix~K of~\cite{migdal2024quantum}. This function involves auxiliary functions $A(\Delta)$, $B(\Delta)$, $C(\Delta)$ defined in earlier appendices of that paper. These functions were precomputed in \Mathematica\ and stored as fifth-order interpolations through a dense table of results obtained by high-precision numerical integration.

The poles of $V(p)$ yield the spectrum of intermittency exponents, which includes both rational values and complex values determined by the zeros of $\zeta(s)$. The entire function $f(-1-p)$ does not contribute to this spectrum of poles; it affects only the residues. This is why the spectrum reduces to rational numbers and the zeros of the zeta function---the basic objects of number theory. The poles $p_n$ with positive real part, relevant for the small-distance expansion of the correlation function, are independent of the space dimension (see Table \ref{tab:poles}).

Complex poles lead to subleading oscillations on a logarithmic scale~\cite{migdal2026Zeta}. Complete details of the analytical and numerical computations involved in this project are summarized in a collection of \Mathematica\ notebooks~\cite{MB40, MB41, MB42, MB43, MB44, MB45, MB46}, which can be freely downloaded to verify these computations.

\end{document}